\documentclass[iop]{emulateapj}

\usepackage{amsmath}
\usepackage{amssymb}
\usepackage{latexsym}
\usepackage{graphicx}
\usepackage{natbib}
\usepackage{color}

\newcommand{\io}{I_1}
\newcommand{\itw}{I_2}
\newcommand{\ith}{I_3}

 \shorttitle{Coronal velocity fluctuations}
 \shortauthors{Morton et al.}

\begin{document}

\title{A global view of velocity fluctuations in the corona below 1.3 $R_\odot$ with CoMP}
\author{R. J. Morton$^{1,2}$, S. Tomczyk$^{2}$ \& R. F. Pinto$^{3,4}$ }
\affil{$^{1}$Department of Mathematics \& Information Sciences, Northumbria University, Newcastle Upon Tyne,
NE1 8ST, UK\\
$^{2}$High Altitude Observatory, National Center for Atmospheric Research Boulder, CO, USA\\
$^{3}$Universit\'e de Toulouse, UPS-OMP, IRAP, Toulouse, France\\
$^{4}$CNRS, IRAP, 9 Av. colonel Roche, BP 44346, F-31028 Toulouse cedex 4, France
}
\email{richard.morton@northumbria.ac.uk}

\begin{abstract}
{The Coronal Multi-channel Polarimeter (CoMP) has previously demonstrated the presence of Doppler velocity fluctuations in the solar corona. The observed 
fluctuations are thought to be transverse waves, i.e. highly incompressible motions whose restoring force is dominated by the magnetic tension, some of which demonstrate clear periodicity. We aim 
to exploit CoMP's ability to provide high cadence observations of the off-limb corona to investigate the properties of velocity fluctuations in a range of coronal features, providing insight into how(if) the 
properties of the waves are influenced by the varying magnetic topology in active regions, quiet Sun and open fields regions.
An analysis of Doppler velocity time-series of the solar corona from the $10,747$~{\AA} Iron XIII line is performed, determining the velocity power spectra and using it as a tool to probe wave 
behaviour. Further, the average phase speed and density for each region are estimated and used to compute the spectra for energy density and energy flux. In addition, we assess the noise levels 
associated with the CoMP data, deriving analytic formulae for the uncertainty on Doppler velocity measurements and providing a comparison by estimating the noise from the data. It is found that the entire corona is replete with transverse wave behaviour. The corresponding power spectra indicates that the observed velocity fluctuations are predominately generated by 
stochastic processes, with the spectral slope of the power varying between the different magnetic regions. Most strikingly, all power spectra reveal the presence of enhanced power occurring at $
\sim3$~mHz, potentially implying that the excitation of coronal transverse waves by $p$-modes is a global phenomenon.}
\end{abstract}



\section{Introduction}
There is currently significant interest in MHD waves in the solar atmosphere and whether they transport enough energy to play a significant role in solar atmospheric heating and the acceleration of 
the solar wind (see, e.g., the following reviews and references within, \citealp{NARULM1996}, 
\citealp{KLI2006}, \citealp{ERDBAL2007}, \citealp{MATVEL2011}, \citealp{PARDEM2012}, \citealp{CRA2012}, \citealp{HANVEL2012}). Enthusiasm for 
wave-based theories of heating and acceleration has been renewed in recent years, with observations suggesting the presence of transverse waves in many 
distinct plasma structures defined by the magnetic field in both the corona and the chromosphere (e.g., \citealp{DEPetal2007}, \citealp{OKAetal2007}, 
\citealp{TOMetal2007}, \citealp{ERDTAR2008}, \citealp{MORetal2012c, MORetal2013}, \citealp{PERetal2012}, \citealp{HILetal2013}, \citealp{NISetal2013}, 
\citealp{THUetal2014}).

{In complex magnetised plasmas, such as the solar chromosphere and corona, MHD waves
cannot typically be characterised as purely Alfv\'en or purely fast modes, but have
mixed properties (although there are exceptions such as the  $n=0$ torsional Alfv\'en 
wave, e.g., \citealp{SPR1982}). The kink wave is one example of a transverse wave mode with mixed properties (see, e.g., \citealp{SPR1982}, \citealp{EDWROB1983}, \citealp{GOOetal2009,GOOetal2012}).
The properties of the kink wave vary depending on the ratio of the 
wavelength ($\lambda$) to the radius ($r$) of the magnetic flux tube that supports the wave. In the so called Thin Tube (TT) limit, when 
$r<< \lambda$, the kink wave has the following properties: \textit{(i)} high incompressibility; \textit{(ii)} the ability to transport vorticity; \textit{(iii)} the dominance of magnetic 
tension as the restoring force (\citealp{GOOetal2009,GOOetal2012}).}

Current 
instrumentation indicates that the chromosphere and corona is finely structured, with typical transverse scales of $\sim400$~km (\citealp{DEPetal2007c}, \citealp{MORetal2012c}, 
\citealp{BROetal2013}, \citealp{MORMCL2013, MORMCL2014}, \citealp{WINetal2014}). The fine-scale structure is seen to support MHD waves, and 
theoretical considerations demonstrate that these (likely) over dense magnetic flux tubes act as waveguides, funnelling the wave energy through the solar 
atmosphere. The propagation speeds of observed kink waves are typically in excess of 50~km\,s$^{-1}$ in the chromosphere (\citealp{JESetal2015}) and 
$200$~km\,s$^{-1}$ in the corona (\citealp{TOMetal2007}; \citealp{MORetal2015}). Further, periods have been observed as short as 40~s (\citealp{HEetal2009}, 
\citealp{MORMCL2013}), although periods of 100~s-500~s are more usual with current observational capabilities. Hence, for a somewhat typical case
with $r=400$~km, $c_p=100$~km\,s$^{-1}$ and $P=100$~s, the ratio $r/\lambda=r/(c_{p}P)\sim 0.04$. This would imply that the currently observable 
kink modes lie in the TT regime where the wave displays their incompressible qualities.

 Previous observations of transverse waves have largely been through imaging observations, revealing the ubiquity of kink motions characterised by the non-axisymmetric displacement of flux tubes. These observations have been 
ideal for establishing the
existence of such wave modes and also providing estimates for their typical properties (amplitudes, periods) in the different magnetic environments. 
\cite{MCIetal2011} demonstrated that the transverse waves are found in active regions, the quiet Sun and coronal holes, and further 
suggested that typical amplitudes are greater in the coronal holes and smaller in active regions, with periods between 100-500~s. Similar results are found for chromospheric features 
(\citealp{PERetal2012}, \citealp{MORetal2013b}, \citealp{JESetal2015}). 

Even after numerous observations of these waves, major uncertainties about their contribution to energy transfer still remain. For example, it is still unclear exactly what role the observed waves play 
(if any) in heating the solar atmosphere. While estimates for the energy content (and flux) of the waves have been given, the values are still subject to a great deal of uncertainty from both 
observational  (e.g., \citealp{VANetal2008c}, \citealp{MCIetal2011}, \citealp{MORMCL2013}, \citealp{NISetal2013}, \citealp{THUetal2014}) 
and theoretical standpoints (\citealp{GOOetal2013}, \citealp{VANetal2014}).  

Now that the presence of the transverse waves in the solar atmosphere has been established, the focus 
of wave observations should shift to measuring other attributes that can be used for testing current wave-based theories.  Although, this may be an onerous task with imaging observations. The 
typical process of measurement, at present, is relatively cumbersome and time-intensive, however, it can provide well-constrained measurements. Some progress has been made towards this goal 
though. For example, \cite{HILetal2013} and \cite{MORetal2013b} derived velocity power spectra for kink waves in prominences and fibrils 
respectively, allowing for an initial comparison to the velocity power spectra derived from motions of granules and 
magnetic bright points. The results suggested an apparent correlation between the spectra hinting the waves may be driven by the photospheric motions (see also, 
\citealp{STAetal2013,STAetal2014, STAetal2015}). The 
generation of incompressible waves by the turbulent convective motions in the photosphere has been a long held belief (e.g., \citealp{OST1961}) and typically is the driving mechanism for transverse
waves in models of heating and wind acceleration.

\medskip

Observations that allow for Doppler diagnostics, such as those with the \textit{Coronal Multi-Channel Polarimeter} (CoMP - \citealp{TOMetal2008})
provide a significant advantage over imaging observations in the fact that it is less arduous to generate power spectra. The sampling of particular 
spectral lines allows the measurement of both Doppler velocities and Doppler widths, providing time-series of these quantities. CoMP, at present, is 
unique in providing a global view of the corona with the required spectral resolution to provide high cadence time-series of velocity fluctuations. CoMP data has been used previously to provide a focused look at kink wave propagation in individual features e.g., \cite{TOMMCI2009}; \cite{THRetal2013}, \cite{DEMetal2014}.

One of the interesting features observed in the CoMP data is the appearance of an enhancement of power, centred on 3~mHz, found in a quiescent loop and an open field region (\citealp{TOMetal2007}, \citealp{MORetal2015}). The coronal magnetic fields are generally considered to be rooted in kilo-gauss faculae that form the network and plage regions in the photosphere (\citealp{GAB1976}, \citealp{DOWetal1986}, \citealp{PET2001}), possibly apart from active region features which can emanate from pores and sunspots. Studies of the interaction and scattering of acoustic waves with flux tubes suggest the concentrations of photospheric magnetic flux provide waveguides for \textit{p}-modes to leak out from the interior into the lower solar atmosphere (\citealp{SCHCAL2006}, \citealp{JAIetal2011}, \citealp{GASetal2014}). The magneto-acoustic waves can propagate higher into the solar atmosphere, eventually reaching a canopy where the Alfv\'en speed equals the sound speed. At this canopy mode coupling occurs and 
the magneto-acoustic energy is split between both slow and fast magneto-acoustic waves, with the proportion dependent upon the angle of the magnetic
field, e.g.,  \cite{BOGetal2003}, \cite{KHOCOL2006}, \cite{KHOetal2008}, \cite{VIGetal2009},
\cite{FEDetal2009, FEDetal2011}. While, slow modes above this canopy are expected to steepen and shock due to the rise in temperature in the 
upper chromosphere (as evidenced by the dynamics of type-I spicules, e.g., \citealp{DEPetal2004}), the fast magneto-acoustic waves can be reflected due to the 
steep gradient in Alfv\'en speed at the transition region. Theoretical and numerical modelling of wave propagation in a simplified atmosphere have shown, under certain conditions, there is a coupling of the fast wave to the Alfv\'en wave, enabling wave 
energy to cross the transition region and propagate into the corona as Alfv\'en waves. (\citealp{CALGOO2008}, \citealp{CALHAN2011}, 
\citealp{CAL2011}, \citealp{KHOCAL2012}, \citealp{HANCAL2012}). This process by its very nature must generate vorticity, which has to propagated 
through an inhomogeneous corona implying, in principle, that it could excite kink waves. Using a relatively simple model
atmosphere, \cite{HANCAL2012} estimate that a sufficient amount of energy can be
converted from \textit{p}-modes to coronal Alfv\'en waves to explain the estimated energy content of observed kink motions in the corona. Due 
to the ubiquity of small-scale magnetic features across the solar surface and the global nature of $p$-modes, it may be expected that this 
phenomenon is widespread through the atmosphere. 

In the following, CoMP data is utilised to investigate the properties of velocity fluctuations globally in the corona.  The main diagnostic to achieve this is the velocity power spectrum, which is derived for typical coronal regions, i.e., quiet Sun, active, and open field regions. Each region is found to have spectra with qualitatively similar properties, with steep spectral slopes and power enhancements at $3$~mHz. It is evident that the power enhancement is present throughout the corona and confirms its global nature. However, each spectra has distinct power laws and the magnitude of the power also varies between the regions. The measured flux of wave energy density and flux are found to be inhomogeneous through the corona, with the greatest flux energy in the quiescent regions. To complement the analysis, we derive analytic formula for the errors related to the analytic fits used for the CoMP data products, which enables us to assess the limitations of the CoMP observations.

\begin{figure*}[!tp]
\centering
\includegraphics[scale=0.51, clip=true, viewport=2.cm 7.4cm 18.cm 23.cm]{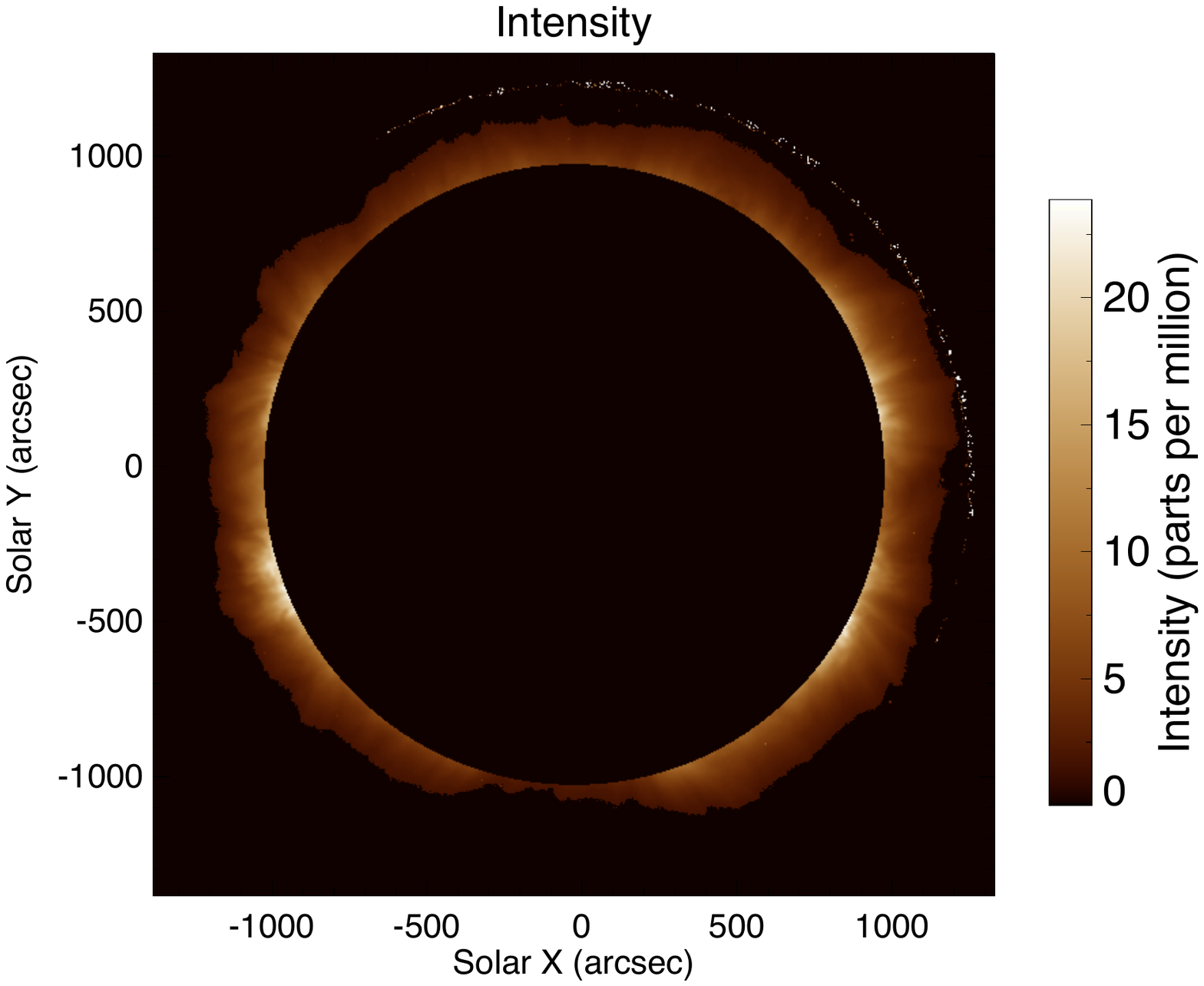}
\includegraphics[scale=0.6, clip=true, viewport=2.cm -0.3cm 15.cm 13.cm]{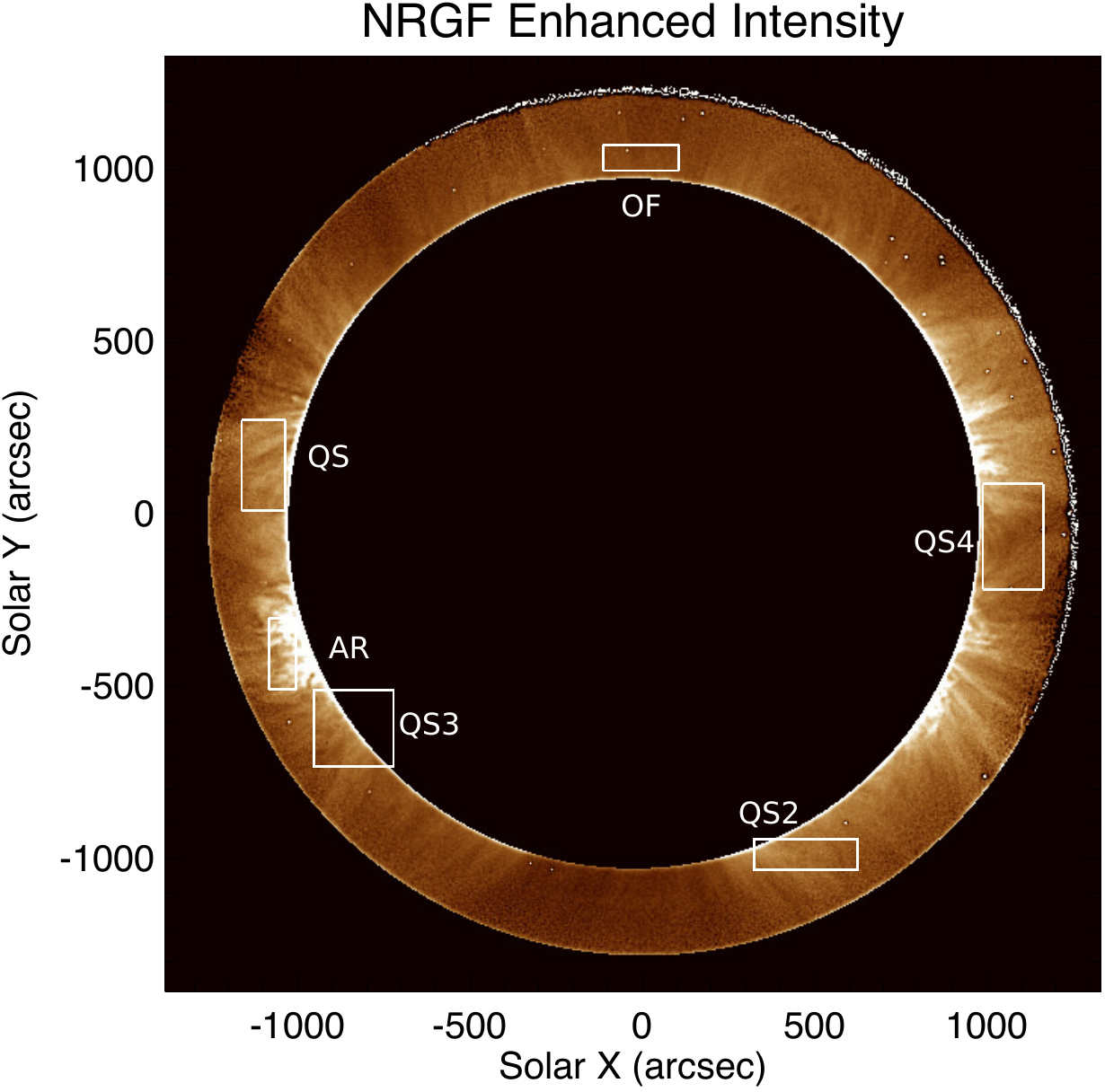}  
\caption{The CoMP field of view on the 27 March 2012 at 18:51:02~UT. The left hand panel displays the line centre intensity, which
is scaled in millionths of the solar disk intensity. The right hand panel shows the intensity enhanced by a NRGF filter, revealing some
of the fine scale structure in the corona. The pixels used to obtain the average power spectra (see, Figure~\ref{fig:av_spec}) are highlighted by the boxes,
for typical coronal features, i.e. an open field region (OF), the quiet Sun (QS) and active region (AR). An additional three quiet Sun region (QS 2-4) are examined in Figure~\ref{fig:av_spec2}. }\label{fig:comp_fov}
\end{figure*}

\section{Observations and data reduction}
The data used here were obtained with the CoMP on the
27 March 2012 at {18:51:02}~UT to {20:13:02}~UT. The details of the acquisition and reduction of CoMP data are fully described in \cite{TOMetal2008}
and we make use of the final data product that has a cadence of $30$~s and a pixel size of $4".46$. We note the individual frames in the 
final data product are produced by averaging over 16 consecutive images. The final data products
are intensity images of the corona at three wavelengths (10745.0~{\AA} - $\io$, 10746.2~{\AA} - $\itw$ and 10747.4~{\AA} - $\ith$), which are positions centred on 
the 10747~{\AA} {Fe}{XIII} emission line (peak formation temperature of $\sim1.6$~MK in ionisation equilibrium). Following \cite{TIAetal2013}, for each pixel in 
the CoMP field of view in each time frame we calculate the central intensity, Doppler velocity shift and Doppler width of the line profile using an analytic fit of a 
Gaussian to the intensity values at each wavelength position\footnote{Note, there is a typographical error in the formula given in 
\cite{TIAetal2013} for the Doppler width, missing the square over the $d$.}. The equations for these quantities are
\begin{eqnarray}
v &=&\frac{w^2}{4d}(a-b),\\
w &=&\sqrt{\frac{-2d^2}{a+b}},\\
i &=&\itw\exp{\frac{v^2}{w^2}},
\end{eqnarray}
where $v$ is the Doppler velocity, $w$ is the Doppler width and $i$ is the line centre intensity.
The $a$ and $b$ are functions of $\io, \itw, \ith$, namely,
\begin{equation}
a=\ln{\frac{\ith}{\itw}},\qquad b=\ln{\frac{\io}{\itw}}
\end{equation}
 and $d$ is the spectral step size. For the Doppler velocity time-series, the solar rotation is removed 
in the manner suggested in \cite{TIAetal2013}.

This is the same data set used in \cite{BETetal2014}. It consists of 164 images, almost uninterrupted. One image at 19:59:33~UT was of low quality
so was replaced by an image interpolated from the neighbouring images in the time-series. An example image from the data set showing the CoMP field 
of view is shown in Figure~\ref{fig:comp_fov}. In addition, the figure shows an image at 10746.2~{\AA} that has been enhanced with a normalising radial gradient filter (NRGF) to allow some of the coronal structures to be better visualised.

The central intensity images are then aligned using cross-correlation and the same shifts are applied to the Doppler velocity 
images. The results of the cross-correlation suggest that the residual motions of the co-aligned data is less than 0.1 pixels.

\medskip
Further, before the main series of data were taken, a sequence of five-point line scans for the full field of view were taken from {17:44:56}~UT to {18:46:14}~UT, alternating between the {Fe}{XIII} lines at $10747$~{\AA} and $10798$~{\AA}. For each emission line scan, the intensities are fit with a Gaussian, allowing the line core intensity to be estimated. These line core intensities will be used for density diagnostics.

\begin{figure*}[!tp]
\centering
\includegraphics[scale=0.2, clip=true, viewport=0.cm 0.cm 85.5cm 60.2cm]{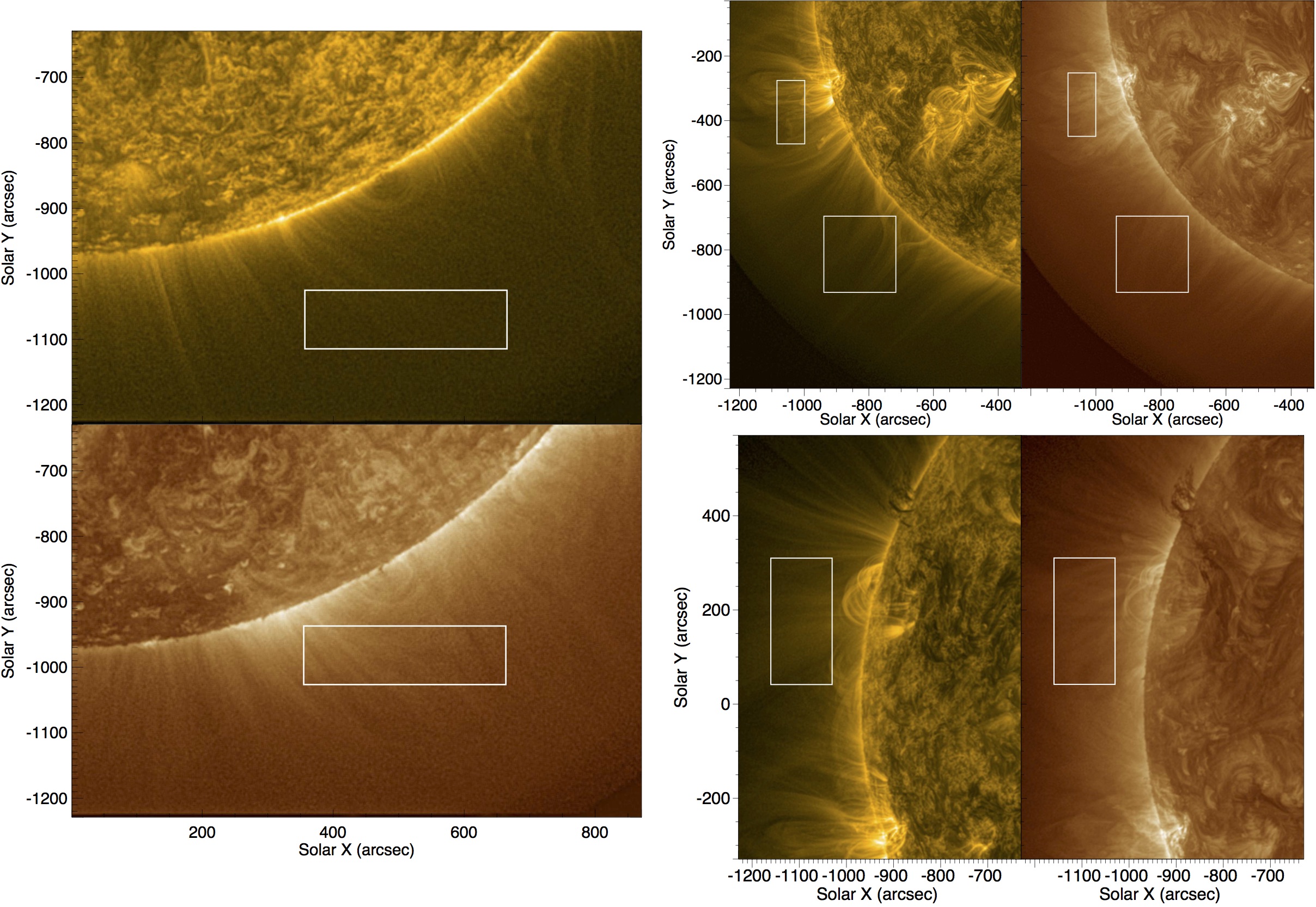}
\caption{High resolution SDO/AIA images showing coronal emission in 171~{\AA} (left panels) and 193~{\AA} (right panels). The images reveal the magnetic topology of the regions analysed with CoMP. The left hand panels show QS2, the top right hand panels shows AR and QS3, the bottom right panels show QS. The images have been subject to an MGN filter (\citealp{MORDRU2014}) to enhance the fine-scale structure.}\label{fig:AIA_fov}
\end{figure*}
 
\medskip

In the following we will investigate velocity fluctuations throughout the corona. 
Regions with varying magnetic geometries and field strengths, i.e., open field region (OF), quiet Sun (QS, QS2, QS3, QS4) and active region (AR)
are selected in CoMP images. The identification of regions is aided with coronal data from the \textit{Solar Dynamics Observatory} (SDO - \citealp{PESetal2012}) \textit{Atmospheric Imaging Assembly} (AIA - \citealp{LEMetal2012}) and magneto-grams from the \textit{Helioseismic and Magnetic Imager} (HMI - \citealp{SCHetal2012}) that reveal the photospheric magnetic flux, which also allow for estimates of the field line connectivity 
using PFSS extrapolations (\citealp{SCHDER2003}). The magneto-grams and extrapolations are not shown. 

\begin{figure}[!tp]
\centering
\includegraphics[scale=0.23, clip=true, viewport=0.cm 0.cm 40.cm 30.5cm]{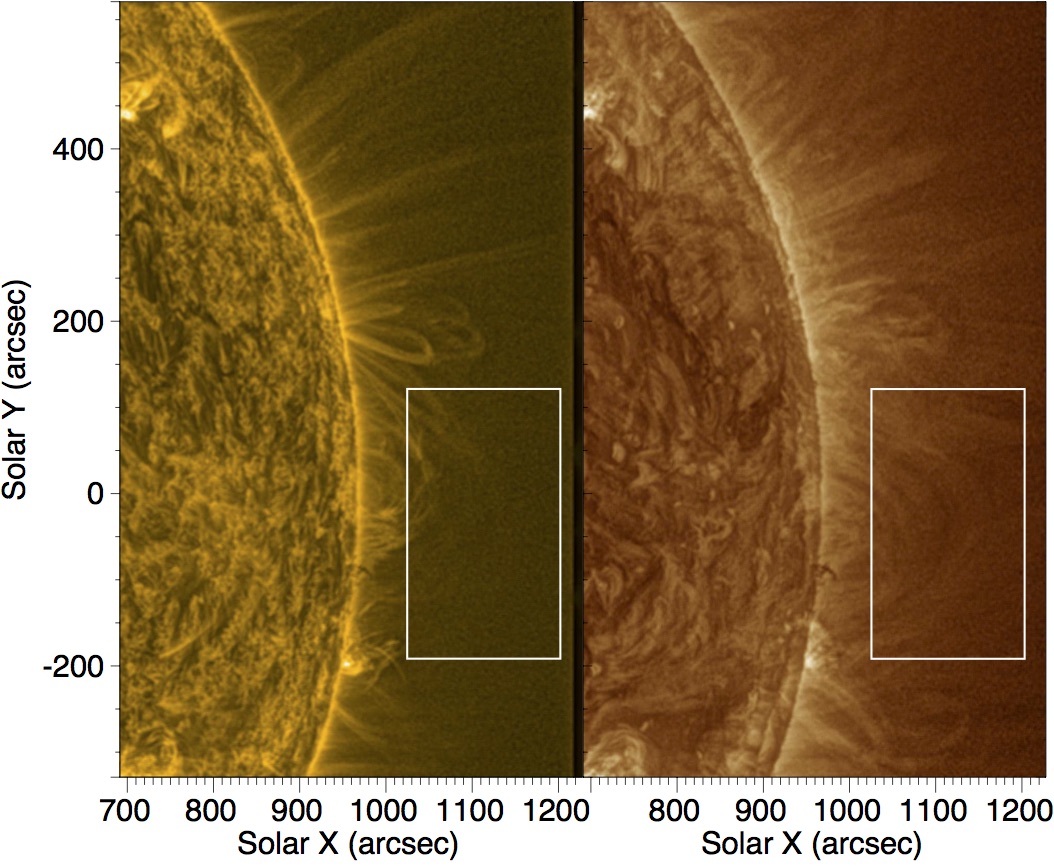}
\caption{Similar to Figure~\ref{fig:AIA_fov} but showing the region QS4.}\label{fig:AIA_fov_2}
\end{figure}

Six regions in the corona are selected for analysis and are chosen as they are representative of active, open field and quiet Sun regions. These regions are highlighted by boxes in 
Figure~\ref{fig:comp_fov}\footnote{ 
Note, some boxes overlap the occulting disk but only pixels that have emission for the entire time-series are analysed.}. In Figures~\ref{fig:AIA_fov} and 
\ref{fig:AIA_fov_2}, data from SDO/AIA is presented that provides a high resolution (0".6/pixel) view of the corona in the 171~{\AA} and 193~{\AA} bandpasses. The 
images reveal the fine-scale magnetic structure of the active region (11448) and the quiet Sun regions that are highlighted by the boxes in Figure~\ref{fig:comp_fov}. 
The open field region is analysed in \cite{MORetal2015} so we will not discuss the details here.

The first panels in Figure~\ref{fig:AIA_fov} show a quiet Sun region (QS2) located near the southern pole, which demonstrates that the local magnetic field is composed of both closed and open magnetic structures (the term open is used to refer to magnetic fields that reach the source surface in the PFSS extrapolation). No large scale magnetic flux features are evident in magneto-grams from the preceding days. 

The second 
panels show the active region (AR 11448) located at the limb, which displays a complicated mass of coronal loop structures. Magneto-grams from the proceeding days show the active region
rotating onto the disk and reveal a bipolar magnetic flux concentration that is predominantly east-west orientated, in which the loops are rooted. This implies that AIA and CoMP are observing 
the loops end on, as opposed to the loops being in the plane of sky. There is also a quiet Sun region (QS3) to the south of the active region with a cusp-like structure. Some of the more northern magnetic field associated with this structure may be rooted in a unipolar region that lies close to the active region, but the southern magnetic field emanates from a magnetically quiet region, similar to that of the first quiet Sun feature. 

The third set of panels in Figure~\ref{fig:AIA_fov} display a third quiet Sun region (QS). The magnetic field in this region is predominantly open, 
although its appearance isn't radial like the fields in the other two quiet Sun regions. The magnetic fields originate from two patches of unipolar network magnetic field with opposite polarity. The PFSS extrapolation of the magnetic field in the corona suggests that the closed loops seen clearly in $171$~{\AA} have a footpoint in each of these patches.

The final region (QS4) is shown in Figure~\ref{fig:AIA_fov_2}. The feature of interest is an arcade of trans-equatorial coronal loops that are almost in the plane of sky. The loops have their southern footpoints located close to a decaying active region (11436) and their northern footpoints close to a patch weak unipolar field. 
There is a region between the two where no distinct large-scale photospheric magnetic structure is visible. The exact location for each set of 
footpoints is difficult to distinguish, with the coronal loops in $193$~{\AA} having higher/lower latitude footpoints (for northern/southern hemisphere footpoints) to the loops visible in $171$~{\AA}. 

\begin{figure*}[!tp]
\centering
\includegraphics[scale=0.9, clip=true, viewport=0.cm 0.0cm 18.cm 13.0cm]{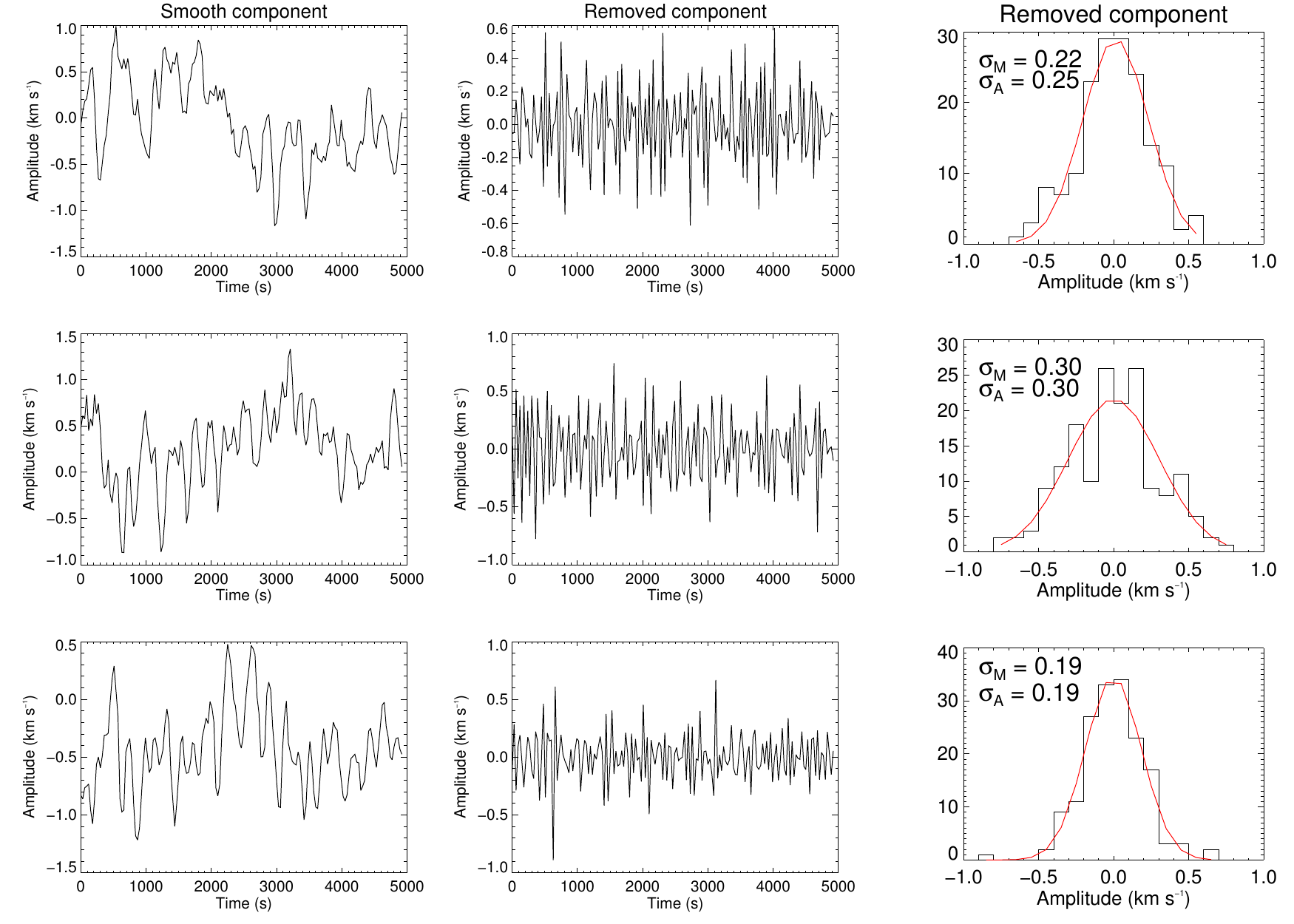}
\caption{Estimating the noise from the data. Three examples of Doppler velocity time-series from single pixels are separated into
the signal (first column) and noise (second column). A histogram of the removed signal is fit with a Gaussian (red line) and the standard
deviation of the measured noise, $\sigma_M$, is given. This can be compared to the analytic estimate for the noise, $\sigma_A$. }\label{fig:signoi}
\end{figure*}

\section{Determining CoMP noise levels}\label{sec:met}
In the following, estimates for the errors on CoMP measurements will be calculated in order to evaluate constraints on results derived here and in future studies. First, analytic formulae are derived to estimate the uncertainties associated with the Doppler velocities, which are then compared to measurements of noise from the data.

\subsection{Uncertainties on the measurable quantities}

The errors on the intensities $\io, \itw, \ith$, used for the analytic Gaussian fit, requires an estimate of the data noise ($\sigma_N$), which is given by
\begin{equation}\label{eq:sigN}
\sigma_N^2={\sigma_p(F)^2+\sigma_d^2+\sigma_f^2+\sigma_{bck}^2(F)+2\sigma_r^2+\sigma_{sd}^2+\sigma_{see}(F)^2},
\end{equation}
where $\sigma_{p}(F)$ is the uncertainty due to photon noise, $\sigma_d$ is the dark current, $\sigma_f$ the  
flat field, $\sigma_{bck}(F)$ is the photospheric continuum background subtraction, $\sigma_r$ the  
readout (factor of two for coronal emission and photospheric background), $\sigma_{sd}$ the digitisation and $\sigma_{see}(F)$ is the seeing noise. The $F$ indicates the noise level is dependent on the intensity flux.

Each time frame of $\io, \itw, \ith$ are a product of averaging over a number of exposures, hence, the data noise can be divided by the square root of the number of exposures, i.e., sixteen. It is likely that the photon noise, the background subtraction and the seeing will dominate the noise. The uncertainties associated with the flat and dark noise are small and can be neglected after the averaging. 
While we are able to confidently provide the uncertainties for photon, background and read noise, it is much more difficult to assess the magnitude of the seeing noise. It is expected that the seeing noise is proportional to the intensity gradient, i.e.,
\begin{equation}\label{eq:seeing}
\sigma_{see}^2=\left(\frac{d I}{d z}\right)^2\sigma_z^2,
\end{equation}
where $d I/dz$ is the spatial derivative of the intensity and $\sigma_z$ is the uncertainty due to the seeing.  With respect to $\sigma_z$, a value of $0.1$ pixels is used, in line with estimates of the residual motions from the cross-correlation (for a discussion see Appendix A).

\smallskip
Having calculated a measure of the uncertainty for the intensity, the uncertainty on the velocity is obtained from the standard error propagation formula
\begin{equation}\label{eq:gen_sum}
\delta X^2=\sum\limits_i\left(\frac{\partial X}{\partial x_i}\delta x_i \right)^2,
\end{equation}
where $X$ is the calculated quantity, $\delta X$ is the associated uncertainty (standard deviation), $x_i$ are the independent quantities and $\delta x_i$ is 
their uncertainty.  

\noindent The Doppler velocity is defined as
\begin{equation}
v=-\frac{d}{2}\frac{a-b}{a+b}=-\frac{d}{2}\frac{\ln{\frac{\ith}{\itw}}-ln{\frac{\io}{\itw}}}{ln{\frac{\ith}{\itw}}+ln{\frac{\io}{\itw}}},
\end{equation}
which can be re-written as 
\begin{equation}\label{eq:dop}
v=-\frac{d}{2}\frac{\ln{{\ith}}-\ln{{\io}}}{-2\ln{\itw}+\ln{\ith}+\ln{\io}}.
\end{equation}
Taking the partial derivatives of Eq.~\ref{eq:dop} with respect to the measured intensities, we obtain
\begin{eqnarray}
\frac{\partial v}{\partial \io}&=&\frac{d}{2\io(-2\ln{\itw}+\ln{\ith}+\ln{\io})}\nonumber\\
&&+\frac{d(\ln{\ith}-\ln{\io})}{2\io(-2\ln{\itw}+\ln{\ith}+\ln{\io})^2} \\
\frac{\partial v}{\partial \itw}&=&-\frac{d(\ln{\ith}-\ln{\io})}{\itw(-2\ln{\itw}\ln{\ith}+\ln{\io})^2} \\
\frac{\partial v}{\partial \ith}&=&-\frac{d}{2\ith(-2\ln{\itw}+\ln{\ith}+\ln{\io})}\nonumber\\
&&+\frac{d(\ln{\ith}-\ln{\io})}{2\ith(-2\ln{\itw}+\ln{\ith}+\ln{\io})^2} 
\end{eqnarray}
These equations can then be substituted into Eq.~\ref{eq:gen_sum} and using the measured values of $\io, \itw, \ith$ and their calculated uncertainties 
(Eq.~\ref{eq:sigN}), an estimate for $\delta v$ can be obtained. 

Note, here we have assumed that the values of $\io, \itw, \ith$ are independent of each other.
The transmission profile of the tunable filter of CoMP has a bandpass of $1.3$~{\AA}, so there is overlap of contributions from the different wavelength positions. However, the images in different filter are taken at different times, which means that measured intensities at each wavelength are independent.

\subsection{Noise levels in the data}
In order to provide a comparison for the analytic uncertainties derived above, an estimate for the noise is calculated from the data.
It has been suggested by \cite{OLS1993} that the best method for determining the distribution of noise is to apply a simple
box-car average filter to the data in order to remove the structure, leaving behind the noise. \cite{STAMUR2006} suggest that exploiting multi-scale
methods, e.g.,  {\`a} trous, may improve upon this simple averaging, although we found little difference for this specific data set. 

For the CoMP data, a box-car smoothing function of length three is applied to the time-series for each pixel in the data set. The smoothed filtered series is then subtracted from the original signal to leave the estimate for the noise. Figure~\ref{fig:signoi} shows three randomly selected time-series showing the separated filtered and noise parts of the signal. The noisy signals are tested for normality using the Kolmogorov-Smirnov test at the $5\%$ level, taking into account the correction required for unknown mean and variance (e.g., \citealp{LIL1967}). It is found that there is no evidence to reject the null hypothesis, i.e., the data is from a normal distribution, for $\sim96\%$ of the noise signals. The fact that $\sim4\%$ are rejected is in line with the expected Type-I error, so it is safe to assume that all noise signals are very close to being normally distributed (e.g., Figure~\ref{fig:signoi} right panels), as should be anticipated for white noise. Possibly, it may be expected the noise should be a combination of Poisson and Gaussian distributions on the noise, with the Poisson contribution coming from the photons. However, for a large mean value the central limit theorem states that the Poisson distribution tends towards a normal distribution. CoMP images typically contain $>2000$ photons per pixel, hence we should expect the photon noise to be normally distributed.

\begin{figure}[!tp]
\centering
\includegraphics[scale=0.7, clip=true, viewport=2.5cm 0.0cm 18.cm 12.5cm]{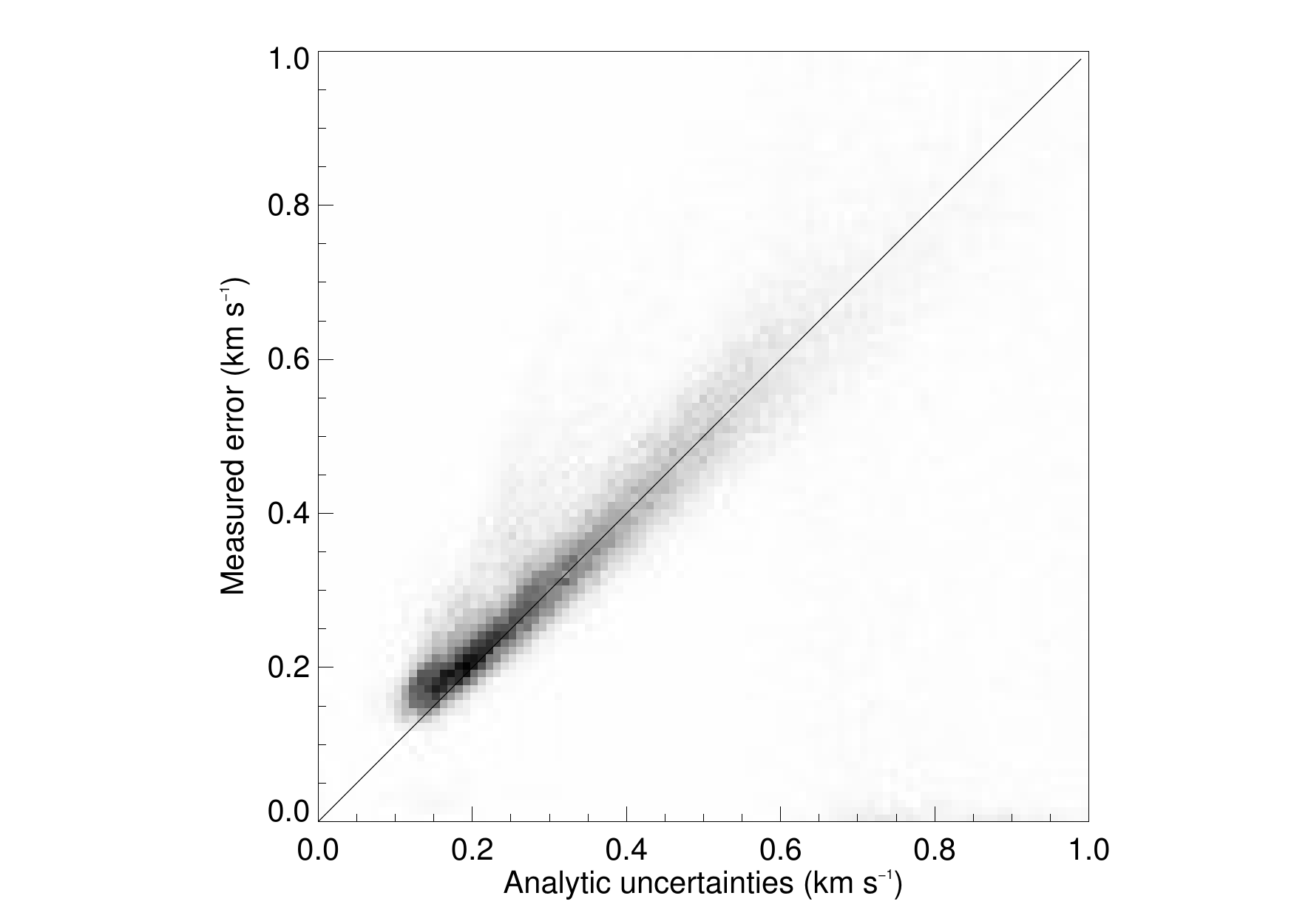}
\caption{A comparison of analytic uncertainties on the Doppler velocity measurements to the root mean square values of noise calculated from the data. The solid line highlights the line of gradient one to facilitate comparison.}\label{fig:meas_vs_estim}
\end{figure}

The standard deviation (root mean square) of the noise signal is calculated and compared to the analytic uncertainties (Figure~\ref{fig:meas_vs_estim}), with the two measures giving comparable estimates.

\section{Global properties of waves in the corona}
Our interest lies in the temporal variation of the Doppler velocity and we will use relatively standard tools of analysis
based around Fast Fourier Transforms (FFT). To begin with, periodograms are obtained for the typical velocity power spectrum for 
each of the boxed regions shown in Figure~\ref{fig:comp_fov}. To reduce spectral leakage in the frequency bins, initial processing steps are performed. First, the mean value of the Doppler velocity for the time-series is subtracted. This corresponds to suppressing the DC 
component of the time-series in the resulting power spectra. Next, the time-series is subject to apodisation with a Hann function.

Now, the following is applied to each boxed region. To find the average power in a region and its variance, the power from each time-series in a particular frequency bin is combined to provide a probability distribution function (PDF) for the velocity power at a certain frequency, from which 
a weighted sample mean, $\mu$, and the uncertainties on the estimate of the mean, $\sigma$, are obtained. The power in each frequency bin has a PDF that appears to be log-normal distributed. Hence, the mean and standard error of the distribution are calculated by binning the natural log of the power, where Poisson errors are used for weighting the calculation of the means that correspond to the uncertainties associated with the power binning. 
A linear function is fit to the means, weighted by the $\sigma$'s \citep{MAR2009}, in log-log space, such that when we transform back to velocity power this corresponds to a fit to the function $10^af^b$, where $a$ and $b$ are determined from the fit parameters. The fit is performed in two separate regimes, for frequencies from 0.2-2.0~mHz and 4.1-11~mHz due to obvious presence of oscillatory power around $3$~mHz.

Before discussing the results, it is worth highlighting that the Doppler velocities from CoMP are a measure of the averaged value of velocities of many over-dense magnetic field lines contained within a single pixel. This naturally leads to smaller measured values for the Doppler velocities (\citealp{DEMPAS2012}; \citealp{MCIDEP2012}). Hence, the magnitude of the power spectra measured here systematically underestimates the magnitude of the wave power and any other quantities estimated from the velocities. The number of unresolved magnetic field lines within a single pixel may play an important role in modifying the measurable Doppler velocity (\citealp{MCIDEP2012}). There is the possibility that the number of over-dense magnetic field lines will vary from active regions to open field regions/coronal holes, which could influence observed variations in the velocity amplitudes from region to region. However, we are not aware of any studies on the populations of magnetic structures, therefore, we assume that the differences in numbers are small, hence, the relative variations in the magnitudes of power spectra between the magnetically different regions are assumed to be predominantly due to differences in wave behaviour.

\begin{figure}[!tp]
\centering
\includegraphics[scale=0.5, clip=true, viewport=0.cm 0.0cm 18.cm 12.5cm]{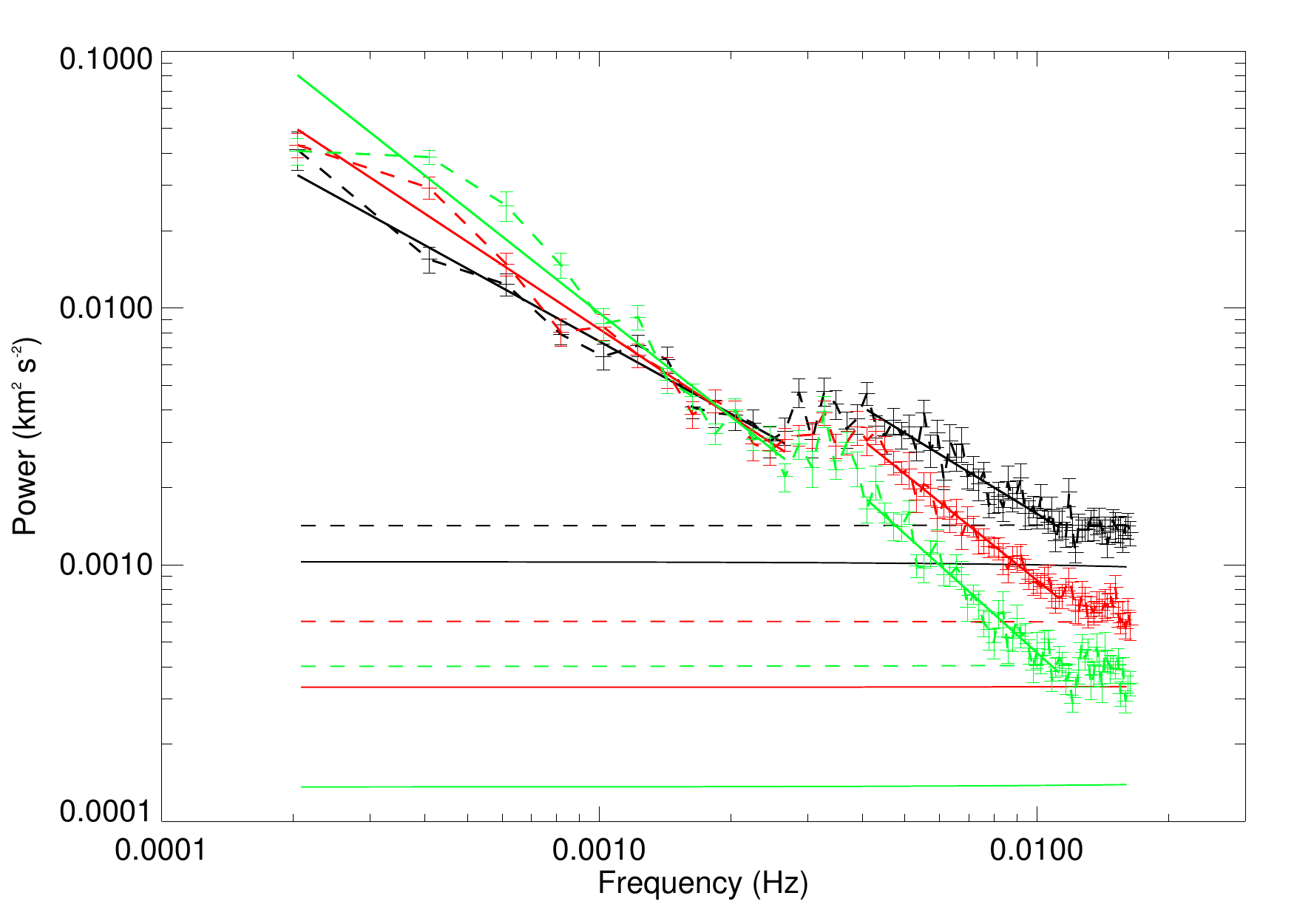} 
\caption{The spatially averaged velocity power spectra. The black data points corresponds the open field region,
the red data points are the quiet Sun region and the green data points are the active region. Each of the sets of data points are 
connected by a dashed line of the same colour. Over plotted are the fitted power law profiles (solid lines), with the values for the fits
given in Table~\ref{tab:meas}. The figure also shows estimates for the noise levels without seeing uncertainties (horizontal solid lines) and with them
(horizontal dashed lines). }\label{fig:av_spec}
\end{figure}

\begin{figure}[!tp]
\centering
\includegraphics[scale=0.5, clip=true, viewport=0.cm 0.0cm 18.cm 12.5cm]{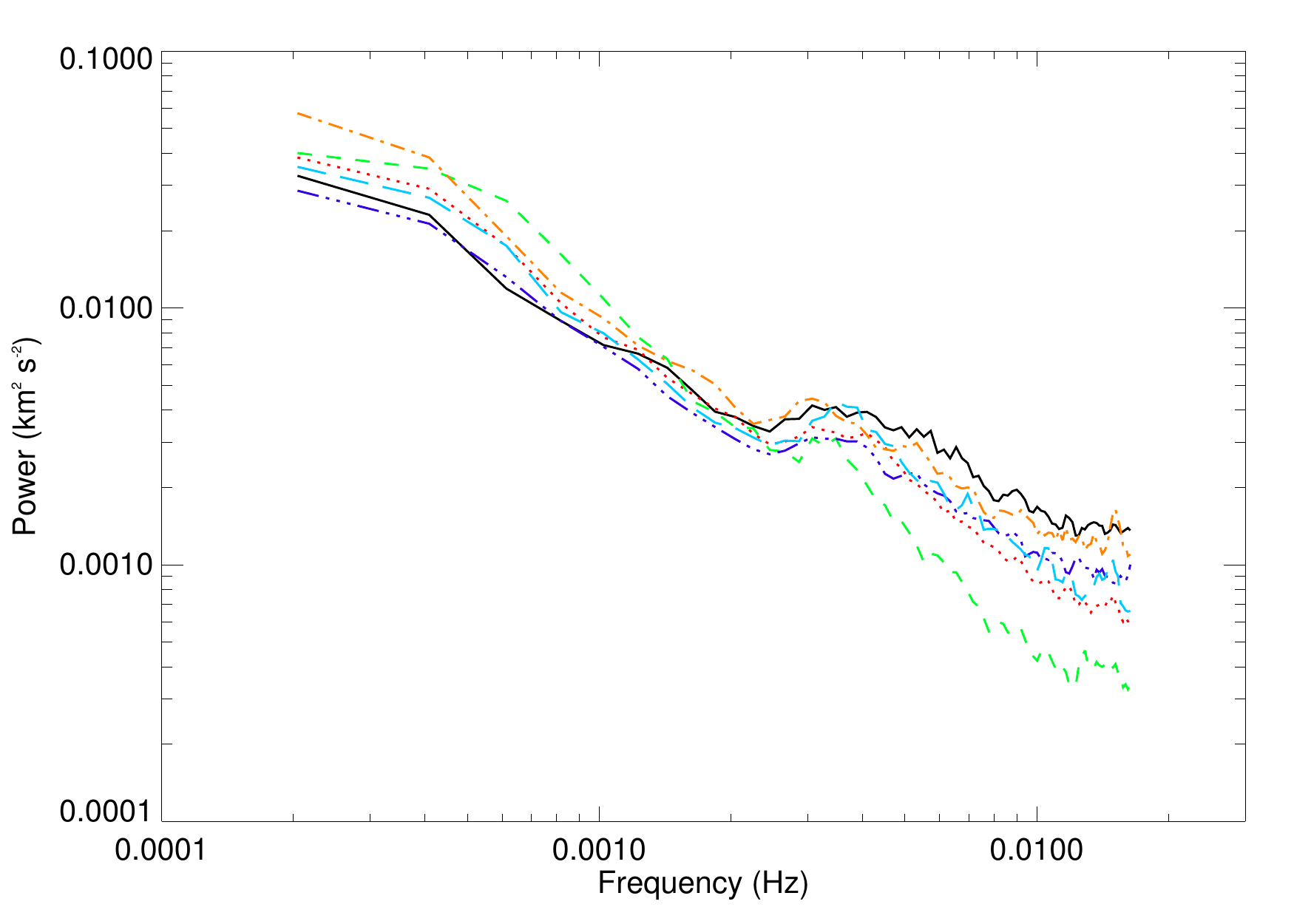} 
\caption{The spatially averaged velocity power spectra including the additional regions. The figure is similar to that shown in Figure~\ref{fig:av_spec},
where: black (solid) line OF; red (dot) line QS; green (dash) line AR; orange (dash-dot) line QS2; blue (long dash) line (QS3);
purple (dash triple dot) line (QS4). Note the spectra have been smoothed with a 3-point box-car function for clarity. The variance and uncertainties for the 
QS2-4 features is comparable to those shown in Figure~\ref{fig:av_spec}. }\label{fig:av_spec2}
\end{figure}

\subsection{Average wave power in the corona}\label{sec:av_pow}
The calculated power spectra for three of the identified regions are shown in Figure~\ref{fig:av_spec} (AR, QS, OF) and all regions are shown in Figure~\ref{fig:av_spec2}. 
It is immediately clear that each power spectra has a slope that is $\propto f^{-b}$, i.e., dominated by an underlying power law spectrum, and that there are 
significant differences between the three spectra. The results of fitting the slopes (Table~\ref{tab:meas}) elucidate these features, showing that the steepness
of the slopes increases from the OF, to QS, to AR. The spectra demonstrate that for the the low frequencies, $f<2$~mHz, the power is initially greater in 
AR, with the OF and QS showing similar values for power (i.e., within 2-3 $\sigma$).
The power spectra then converge around 1~mHz, with each spectrum showing a significant increase in power around 
3~mHz. This enhancement of power provides a break in the $\propto f^{-b}$ trend, hence, is the reason we choose to break up the fitting of the spectra 
into two regimes. This increase in power around 3~mHz was previously identified in the large quiet Sun loops studied in \cite{TOMMCI2009} and the open field region in \cite{MORetal2015}. These 
results suggest that this feature isn't restricted to velocity fluctuations in certain magnetic geometries, but is prevalent through out the corona. 

After the enhancement, as frequency continues to increase, the slope of the power spectra reverts back to that of a power law. Interestingly, the different 
regions show visibly different values for the velocity power at the high frequencies, with OF>QS>AR. As the velocity power is proportional to velocity 
amplitude, this indicates a difference in velocity amplitudes between the regions. 

Below $\sim100$~s, a 
knee appears in all the power spectra and they become almost flat. The simultaneity of the flattening suggests that this is due to noise dominating 
the power. The difference in the power of the noise between regions is likely partially due to the variation in emission between them, i.e. $\delta v(I)$. 
The lower intensities in open field regions and coronal holes compared to active regions, for example, means a lower signal to noise, hence, the magnitude of the errors in the estimated Doppler velocity increases. In turn, this leads to a higher noise level in the power spectra. This is confirmed by using the noise estimates to calculate the power of the noise for each spectra (Figure~\ref{fig:av_spec}).

In Figure~\ref{fig:av_spec2}, the power spectra for all highlighted regions are shown. The plotted power spectra have been smoothed with a box-car filter of length three for aesthetic reasons only. The levels of the variance of the power for the additional quiet Sun regions are similar to that shown in the power spectra in Figure~\ref{fig:av_spec}. It is found that the magnitude of the power spectra and also the spectral slopes of the additional regions are consistent with the previous quiet Sun measurements, somewhat bound between the open field and active region measurements.

\begin{figure*}[!tp]
\centering
\includegraphics[scale=1.07, clip=true, viewport=0.5cm 8.5cm 18.cm 12.7cm]{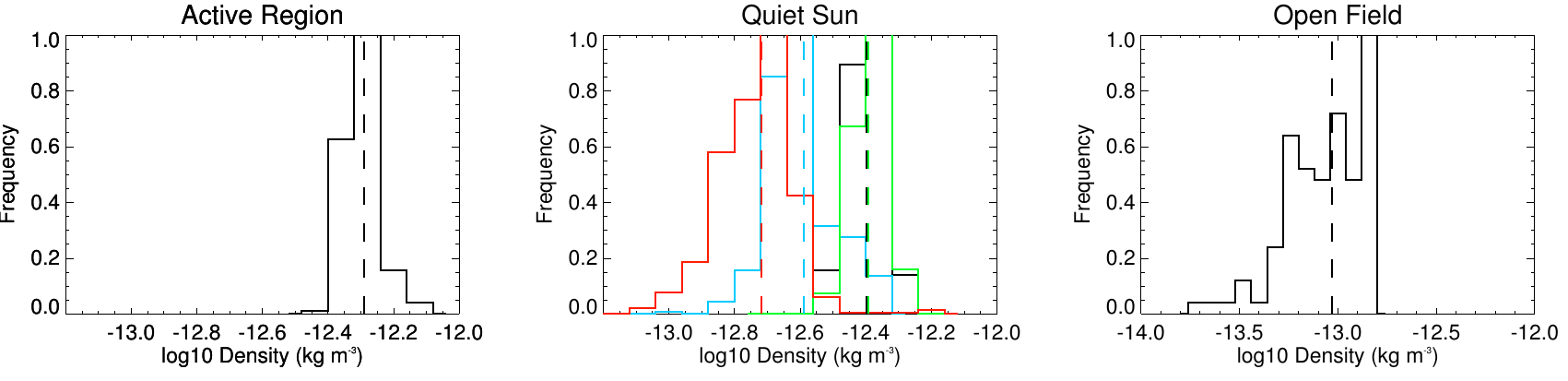}  

\caption{Density measurements from the Fe XII line ratio. The middle panels displays the histograms for QS (black), QS2 (blue), QS3 (green) and QS4 (red). The distributions have been normalised with respect to the largest bin. The dashed lines in each panel mark the mean value
of the density for each distribution.}\label{fig:den_hist}
\end{figure*}

\subsection{Energy density}
Now, we provide an assessment of the relative amount of wave energy stored in the different magnetic geometries by examining the wave
energy density. The energy density is 
$$
\epsilon\propto\rho v^2,
$$
where $\rho$ is taken to be the average mass density of the local plasma. An estimate for the density can be obtained by utilising CoMP measurements of the Fe XIII emission lines at 10798~{\AA} and 10747~{\AA}, the ratio of which is sensitive to the electron number density, $n_e$ (\citealp{FLOPIN1973}). Using the CHIANTI database v7.0 (\citealp{LANetal2012}), electron density versus intensity ratio curves are calculated for a range of heights above the photosphere taking into account the strong influence of photo-excitation on the formation of the two lines. The curves are then used to calculate the electron number density, which is converted to coronal mass density using $\rho=\mu m_pn_e$, where $\mu$ is the mean atomic weight (taken as 1.27 for coronal abundances) and $m_p$ is the proton mass. 

\begin{figure}[!tp]
\centering
\includegraphics[scale=0.5, clip=true, viewport=0.cm 0.0cm 18.cm 13.cm]{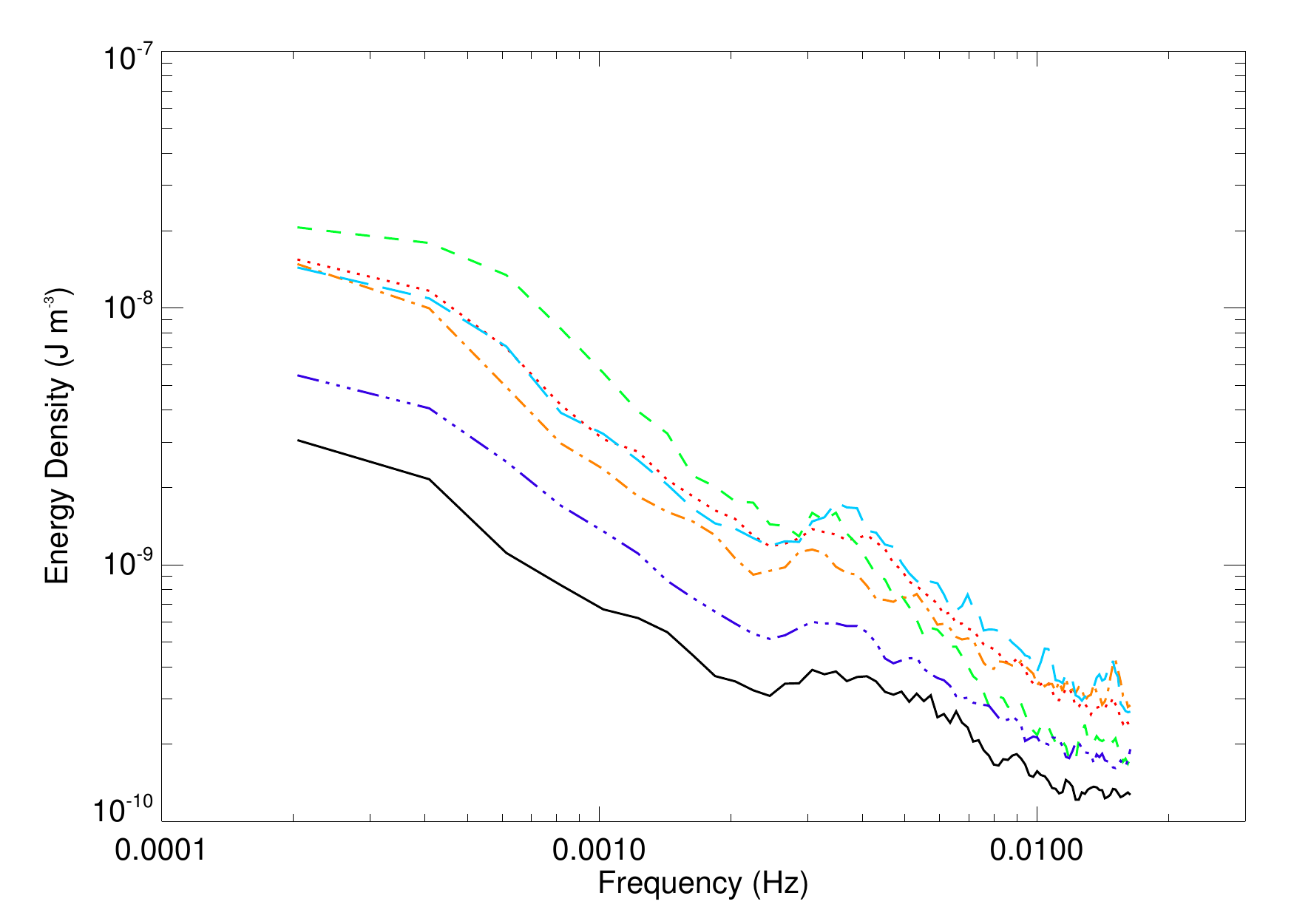}  
\caption{The energy density spectra of the waves in different regions of the corona. The plotted spectra follow the
same colour coding as described in Figure~\ref{fig:av_spec2}.
}\label{fig:en_den}
\end{figure}

In Figure~\ref{fig:den_hist} the distributions of measured density for each region are shown. The power spectra for each region is then multiplied by the corresponding mean density and displayed in Figure~\ref{fig:en_den}.

\subsection{Energy Flux}
\begin{figure*}[!tp]
\centering
\includegraphics[scale=1.07, clip=true, viewport=0.5cm 8.5cm 18.cm 12.7cm]{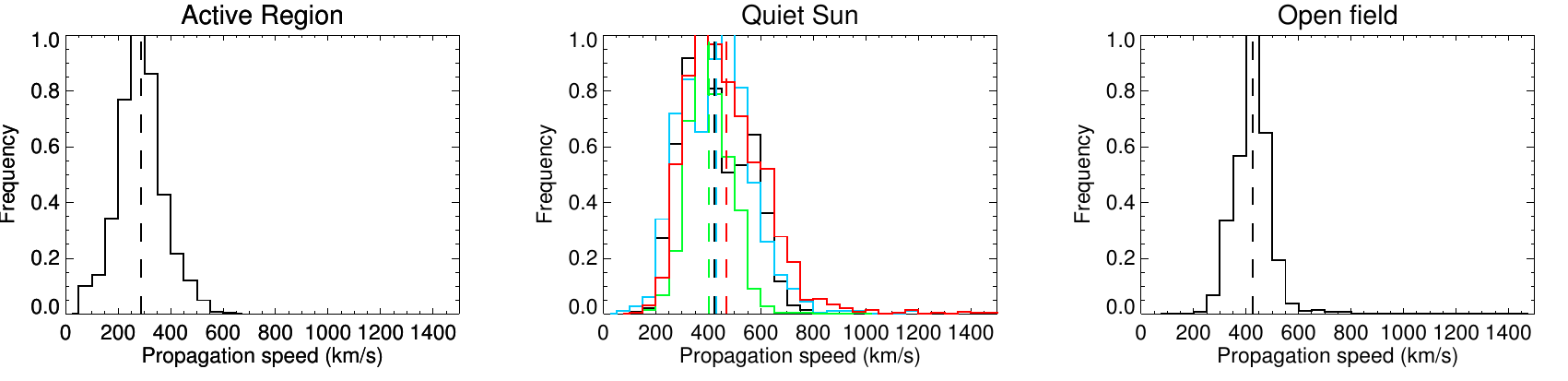}  
\caption{Distribution of propagation speeds for each region. The middle panels displays the histograms for QS (black), QS2 (blue), QS3 (green) and QS4 (red). The distributions have been normalised with respect to the largest bin. The dashed lines in each panel mark the mean value
of the phase speed for each distribution.
}\label{fig:pha_spee}
\end{figure*}

Finally, we also estimate the flux of transverse wave energy through each of the regions. The energy flux is proportional to
$$
F\propto \rho v^2 c_{ph}=\epsilon c_{ph},
$$
where $c_{ph}$ is the phase (or propagation) speed of the wave. An estimate of the propagation speeds of the observed waves can be obtained by cross-correlation of the Doppler velocity signals, full details of the process used are described in \cite{TOMMCI2009} and \cite{MORetal2015}. The distributions of propagation speed measurements in each region are shown in Figure~\ref{fig:pha_spee}. The propagation speeds are significantly smaller in the active region, while the quiet Sun and open field regions display relatively similar values. It is likely that the angle between the magnetic structures that support the waves and the plane of sky will influence these results. The projection of the structure onto the plane of sky leads to a shorter apparent path taken by the wave, hence, measurement would give an underestimation of the propagation speed. This effect may be most influential on the active region measurements, with the AIA data indicating that the loops in the core of the active region are orientated with large angles to the plane of sky. Further, flows along the waveguides will also influence the measured speed of wave propagation. Bearing this in mind, the average value for the propagation speed in each region is determined and combined with the power spectra and density measurements to obtain an estimate for the energy flux for each region.

\section{Discussion and conclusion}\label{sec:discuss}

In the preceding section, time-series of velocity fluctuations in the corona between $1.05-1.3$~$R_\odot$ as observed with CoMP were examined. 
Average power spectra were obtained for various regions of the corona deemed as \lq{typical\rq}, in particular quiet Sun, active, and open field 
regions. {The velocity fluctuations can be associated with transverse waves, most likely the swaying motions of coronal loops, i.e., kink
waves.} Additionally the energy density and 
energy flux of the waves are estimated. In the following, we discuss what inferences can be drawn from these measurements on the 
properties of {kink} waves in the corona.

\begin{table*}
\caption{Measured properties of the different regions. The table displays the results of the $10^af^b$ fit to the power spectra shown in
Figure~\ref{fig:av_spec} in an open field region (OF), quiet Sun regions (QS, QS2, QS3, QS4) and active region (AR). Also given
are the average density, $\rho$, and propagation speed, $c_{ph}$ for the regions and the standard deviations of the distributions.  }
\centering

\begin{tabular}{lccccccc}

\hline\hline\
 & Frequency  &{a} & {b} & {$\chi^2_\nu$} & &$\rho$ & $c_{ph}$\\
& Range (mHz) & & & & & (10$^{-13}$~kg m$^{-3}$) & (km s$^{-1}$)\\[0.2ex]
\hline
\\
OF &0.2-2 &$-4.9\pm0.1$ & $-0.94\pm0.05$ & 1.3 & & 0.9$\pm$0.4 &420$\pm$80  \\
&4-10 & $-4.88\pm0.1$ & $-1.04\pm0.06$ & 1.2 \\
AR & 0.2-2 &$-6.0\pm0.1$ & $-1.34\pm0.04$ & 6.3 & & 5.1$\pm$0.7 &290$\pm$90 \\ 
&4-10 & $-6.4\pm0.1$ & $-1.53\pm0.06$ & 1.5 \\
QS &0.2-2 & $-5.5\pm0.1$ & $-1.13\pm0.04$ & 1.79 & & 4.0$\pm$0.5 &420$\pm$120 \\
 &4-10& $-5.8\pm0.1$ & $-1.37\pm0.04$ & 0.5 \\
QS2 & 0.2-2 &$-5.4\pm0.1$ & $-1.12\pm0.05$ & 2.0 & & 2.6$\pm$0.6 &430$\pm$140 \\
&4-10 & $-4.8\pm0.1$ & $-0.98\pm0.05$ & 1.4 \\
QS3 & 0.2-2 &$-5.4\pm0.1$ & $-1.09\pm0.04$ & 2.1 & & 4.1$\pm$0.6 &400$\pm$80 \\
&4-10 & $-5.5\pm0.1$ & $-1.25\pm0.05$ & 2.6 \\
QS4 & 0.2-2 &$-5.4\pm0.1$ & $-1.07\pm0.04$ & 0.8 & & 1.9$\pm$0.6 &470$\pm$150 \\
&4-10 & $-4.8\pm0.1$ & $-0.95\pm0.05$ & 1.1 \\

\hline

\end{tabular}

\label{tab:meas}
\end{table*}

\subsection{The slope of the spectra}

Each power spectrum measured here demonstrates power law behaviour (except for the enhancement around $3-5$~mHz), potentially indicating that the processes generating the coronal velocity 
power spectra are inherently stochastic in nature. This is perhaps hardly surprising considering that the magnetic fields are rooted in a turbulent photosphere, where the granulation is observed to 
buffet the magnetic flux concentrations (\citealp{BERTIT1996}; 
\citealp{VANBALLetal1998}; \citealp{KEYetal2011}; \citealp{CHITetal2012}). The photospheric velocities imparted on the magnetic field can propagate into the corona via transverse waves, not before a 
significant fraction of the waves are reflected at the transition region, which will also play a role in determining the spectra of the motions that enter into the corona (e.g., \citealp{CRAVAN2005}; \citealp{VERVEL2007}; \citealp{VANBALLetal2011}).

What is also evident is that the slopes for each spectra show a relationship with magnetic geometry. As demonstrated in \cite{MORetal2015}, the power spectra derived for the open field region 
displays a $\sim1/f$ slope, which is also observed in velocity fluctuations in the solar wind (\citealp{BAVetal1982}; \citealp{GOLetal1995}; \citealp{DROB2010}). Further, although not shown, in this 
data set velocity fluctuations in other regions of the corona associated with a predominately open field have an approximately $1/f$ slope. The presence of the $1/f$ slope in the corona may provide 
support for the ideas presented in \cite{VERDetal2012}, who suggest that the spectra observed in the solar wind are already set in the low solar atmosphere, at least in the corona, and advected 
outwards.

The active region power spectra displays the steepest spectral slope with a gradient of $-1.5$, which may be indicative of the development of MHD 
turbulence in the closed coronal loops, particularly Iroshnikov-Kraichnan (IK) type turbulence. Modelling of Alfv\'en waves in closed coronal loops (e.g, 
\citealp{VANBALLetal2011}) demonstrate that Alfv\'en waves injected in from both footpoints leads to counter-propagating wave packets that interact non-linearly and Alfv\'en wave turbulence 
develops and a heating of the coronal plasma ensues. Evidence for such counter propagating transverse waves has been observed previously with CoMP in large-scale quiescent loops 
(\citealp{TOMMCI2009}; \citealp{DEMetal2014}), as well as in an open field region (\citealp{MORetal2015}). However, the slope of the power spectra of fluctuations expected from MHD turbulence is 
still uncertain, with arguments indicating it should be steeper than the $-1.5$ from IK theory (see, e.g., \citealp{BRUCAR2005}; \citealp{PETetal2010}; for reviews on MHD turbulence). 

The quiet Sun power spectra generally fall between these two {\lq extreme\rq} slopes. This may reflect the fact that the quiet Sun regions analysed are composed a mixture of both open and closed 
field regions, as indicated by the SDO data (Figures~\ref{fig:AIA_fov} and \ref{fig:AIA_fov_2} and from PFSS extrapolations), leading to an average spectral profile composed of both the open and 
closed magnetic field contributions. However, the QS4 region appears to contain predominantly closed field but its spectral slope is closer to $-1$, which may indicate other effects play a role in 
determining the quiet Sun spectral slopes.

\subsection{Enhanced power at 3~mHz}
The striking aspect of the spectra is that each one has an enhancement of power around 3~mHz (Fig.~\ref{fig:av_spec2}). 
A convincing theory to explain this feature is the mode conversion from \textit{p}-modes to Alfv\'en waves close to transition region (as discussed in the introduction). As such, \textit{p}-modes would play an important role in generating coronal transverse waves, injecting additional power at $3$~mHz that may the 
cascade to higher frequencies. The conversion of the \emph{global} \textit{p}-modes to transverse waves provides a natural explanation of how such an enhancement is coincident in all the power 
spectra from different regions. 

Due to the enhancement being centred on 3~mHz, it may be tempting to attribute this feature to slow magneto-acoustic waves in the corona, driven by \textit{p}-modes. However, it is known that slow 
waves are strongly damped at much lower heights in the corona than CoMP FOV covers. In particular, slow waves with periods around $\sim3$~mHz above both polar regions (\citealp{GUP2014}, 
\citealp{KRIetal2014}) and 
active regions (\citealp{DEM2009}) appear to be killed off below 30" (1.03~$R_{\odot}$). Although, lower frequency ($<0.02$~mHz) slow waves are 
observed at larger heights ($>$1.06~$R_{\odot}$, e.g., \citealp{OFMetal1997}, \citealp{DEFGRU1998}, \citealp{BANetal2009}, or for 
a review see \citealp{BANetal2011} and references within). Further, for slow waves in a coronal plasma, the perturbation of the plasma velocity is strictly parallel to 
the magnetic field orientation. To 
contribute to the measured power spectra, the slow waves would have to be propagating along magnetic fields parallel or inclined to the line-of-sight. This is odds 
with the close correlation between 
the measured direction of velocity signal propagation in the plane-of-sky and the magnetic field orientation (\citealp{TOMMCI2009}). Further, 
the typical measured propagation speeds are in excess of the estimated local sounds speed and no significant intensity oscillations seen with 
CoMP. As such, it is doubtful that slow waves contribute significantly the the measured velocity signal.

Alternatively, one could suggest that \emph{all} observed transverse waves are excited in the photosphere by the horizontal motions of the convective photosphere and propagate into the corona. The 
results of the spectral evolution of Alfv\'en waves in \cite{CRAVAN2005} demonstrate how an initial driving spectrum of photospheric horizontal motions could potentially lead to an enhancement in 
coronal power spectra due to the reflection at the transition region (e.g., \citealp{HOL1978}). However, it is unclear how active, quiescent and open field regions could produce such a similar 
enhancement of the spectra in this manner. First, photospheric flows are known to be suppressed in regions of increased magnetic flux density (\citealp{TITetal1989}). This would suggest driving 
spectra, and consequently the spectra of the generated transverse waves, would vary depending upon the density of magnetic flux, with evidence for this from chromospheric observations 
(\citealp{MORetal2013b}). Additionally, the magnetic and density scale heights are also likely to vary from region to region, meaning variations in the reflection/transmission profiles. The required 
combination of each of these factors in the different regions to produce a similar enhancement around 3~mHz appears unlikely.

\subsection{Magnitude of the power}
Qualitatively each of the power spectra derived for the different regions have similar characteristic features. However, there are distinct differences between the spectra (Figure~\ref{fig:av_spec} and 
\ref{fig:av_spec2}). An 
apparent anti-correlation exists between the higher frequency power (2.5-10~mHz) and density, with power (hence, velocity amplitude) suppressed in the regions with the greatest density, i.e., the 
power (density) is least (greatest) in the active regions and greatest (least) in the open field regions (Table~\ref{tab:meas}). Similar variations in amplitudes of oscillatory {kink} waves in the corona 
have been inferred from imaging observations with SDO (e.g., \citealp{MCIetal2011}). However, as mentioned, the larger spatial resolution of CoMP leads to an under-resolution of velocity amplitude 
and a direct comparison cannot be provided between the two sets of observations. 

\medskip

The lower frequency part of the power spectra displays different behaviour. There is no apparent correlation between the magnitude of the power in 
the higher frequency part of the spectrum and the 
low frequency part, although this could partially be due to uncertainties associated with the inherently small sampling of velocity amplitudes at 
the lower frequencies. At present it is unclear whether 
the power at these frequencies has an oscillatory component. Long period {kink} waves appear infrequent in SDO imaging data (e.g., 
\citealp{THUetal2014}). The implication would be that 
oscillatory phenomenon do not contribute significantly to the magnitude of power spectra at the lower frequencies. However, this does not rule out 
the observed power is due to transverse waves. 
Waves do not have to be oscillatory in nature, for example they could be pulses. The power at lower frequencies may reflect some of the long term 
evolution the magnetic field undergoes and 
represent the magnitude of the signals that transmit the information about new states of coronal equilibrium (\citealp{DEFetal2014}). For example, 
the recycling time of the coronal magnetic field is 
suggested to take place on time-scales of $\sim1.4$~hours in the quiet Sun (\citealp{CLOetal2004}).

\begin{figure}[!tp]
\centering
\includegraphics[scale=0.5, clip=true, viewport=0.cm 0.0cm 18.cm 13.cm]{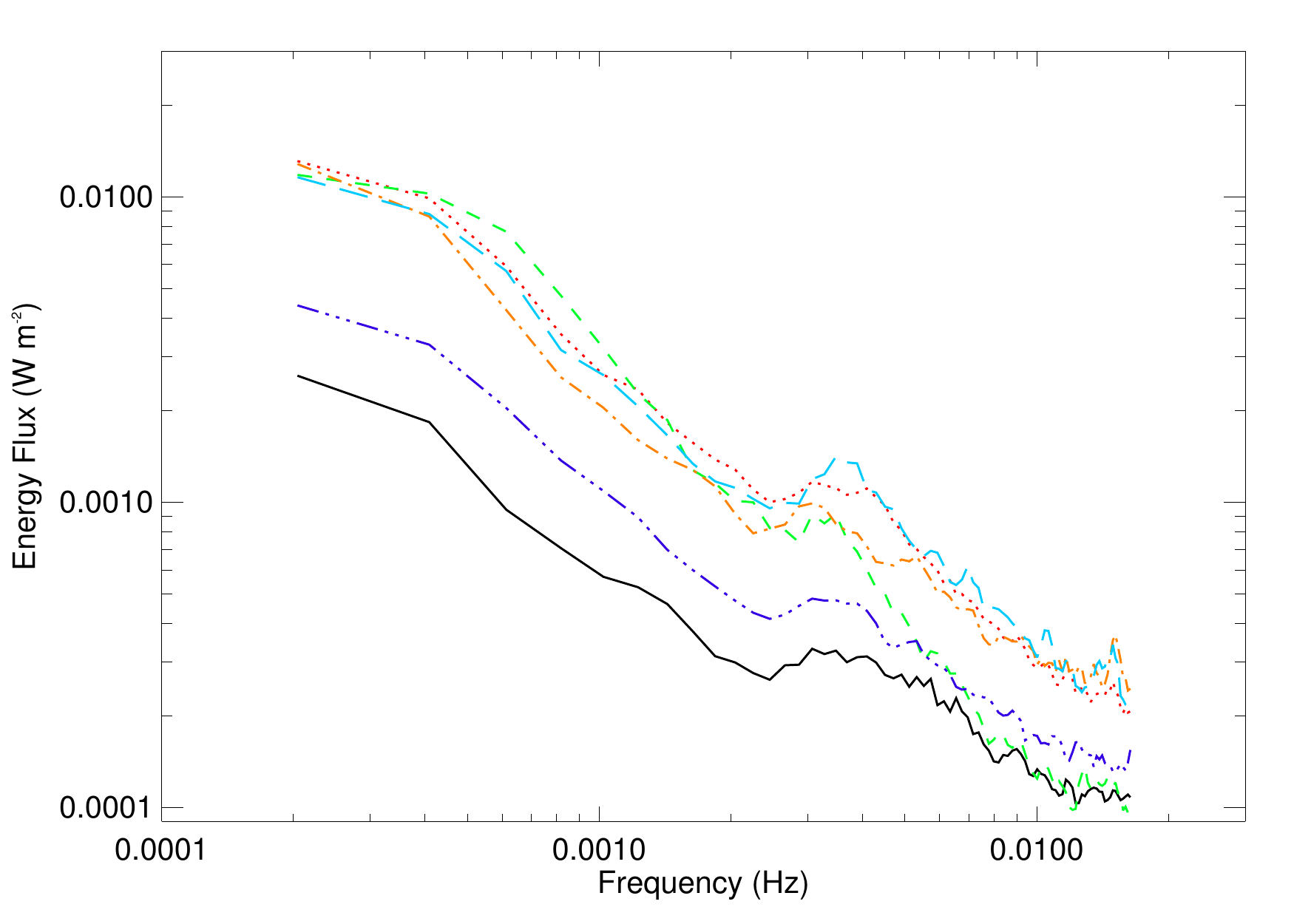}  
\caption{The energy flux spectra of the waves in different regions of the corona. The plotted spectra follow the
same colour coding as described in Figure~\ref{fig:av_spec2}.}\label{fig:en_flux}
\end{figure}

\subsection{Energy budgets}
Finally, the wave energy density and flux for each region were estimated using density and propagation speed measurements from CoMP. It is found that for both quantities, the wave energy content 
across the corona is non-uniform. The energy density reveals that the wave energy stored in the quiet Sun magnetic structures is comparable to that of active regions, both of which exceed that of 
the polar open field region. This is not always the case though, for example the stored wave energy in the large-scale quiescent loops is comparable to the polar open field region.

The smaller values of the polar region energy density will be in part due to the largely one way propagation of waves and low densities. \cite{MORetal2015} 
demonstrated the presence of counter-propagating transverse waves in open field regions, but the amplitude of the inward signals is smaller than the outward signals. In comparison, waves in closed 
structures appear to be excited at both footpoints, hence travel with comparable amplitudes in both directions.

Similar behaviour is also seen in the energy flux estimates. Although the open field region has larger values of propagation speeds compared to the active region (Fig~\ref{fig:pha_spee}, Table~
\ref{tab:meas}), the energy flux remains around a factor of two smaller for frequencies greater than $3$~mHz, increasing to a factor of 3 for lower frequencies. 

Previous estimates of the coronal energy flux of {kink} waves have been provided from SDO observations (\citealp{MCIetal2011}; \citealp{THUetal2014}). By the nature of the motions visible with 
SDO, those observations focus on waves with frequencies between $2-10$~mHz below heights of 1.05~$R_\odot$. The energy flux is calculated for an average value of velocity amplitude across the 
frequencies, however, the estimates imply the energy flux almost uniform throughout the corona. If we calculate the energy flux in a similar manner, we still find a non-uniform energy flux, with the 
active region energy flux one and half times greater than that estimated for the open field regions, and the quiet Sun around a factor of two greater. 

It should be kept in mind that the number of structures contributing to CoMPs velocity signal may vary between the regions, potentially distorting the measure of velocity amplitude to a degree 
(\citealp{MCIDEP2012}). This would have some impact on the relative sizes of energy density and flux, but the number of structures would appear to need to vary by orders of magnitude to make an 
appreciable difference. Hence, we are confident that the measured differences in wave energy densities and fluxes are a physical feature of the corona.

\subsection{Conclusion}
Here, we have analysed the power spectra of velocity fluctuations in the corona in a variety of typical regions with different magnetic geometry.
The fluctuations are thought to be associated with transverse waves ({namely the kink mode}), hence, the spectra give insight into the
typical properties of these waves in the different regions. The results reveal the spectra are qualitatively the same throughout the corona, 
showing a steep spectral slope with a power enhancement around 3~mHz. However, there are distinctions between spectra (i.e., spectral slope; magnitudes of power, energy density, and energy flux), 
implying that the properties of the associated transverse waves vary in the different magnetic geometries. This is perhaps not surprising, and is broadly consistent with the impression from previous imaging observations.

The differing spectral slopes indicate variations in the underlying driving spectrum and evolution of the waves as they propagate through the lower solar atmosphere and corona. The measured 
slopes raise a number of questions, in particular, as to whether the $f^{-1.5}$ dependence in the active region is indicative of MHD turbulence and whether the quiet Sun slopes due to different 
phenomena or a mixture of contributions from features with $f^{-1.5}$ and $f^{-1}$ profiles. 

As highlighted, each spectra shows evidence for a broad peak of enhanced power coincident at the same frequency range. This feature is present in all corona power spectra and indicates that there 
is a common driving mechanism operating on a global scale, that injects significant energy in to the corona around $3$~mHz. The source of the wave energy is potentially from the mode 
conversion of \textit{p}-modes. Such a contribution is neglected in many Alfv\'en wave based heating models but the ubiquity of the feature throughout the corona would imply it plays a major role in 
determining the coronal wave energy budget.  


Moreover, measures of the energy density and flux for each region were estimated, implying that significantly less energy flows through the open field region compared to quiet Sun regions and 
active regions. This is probably due to the largely one way flow of energy along open field lines (\citealp{MORetal2015}), however, varying strengths of the wave driver and fraction of energy reflected 
at the Transition Region in each region will also play a role in determining the coronal energy flux of transverse waves.

\medskip

\acknowledgements
RM is grateful to the Leverhulme Trust for the award of an Early Career 
Fellowship, the Higher Education Funding Council For England and the High Altitude Observatory for financial assistance which enabled 
this work and acknowledges IDL support provided by STFC. RM is also grateful to M. Goossens and G. Verth for discussions on MHD wave theory,
G. Li for enlightening discussions on statistical hypothesis testing, and C. Bethge and H. Tian for assistance with density diagnostics. NCAR is supported by the National Science 
Foundation. RP was supported by the FP7 project \#606692 (HELCATS).

\appendix
\section{Estimating the seeing noise}\label{sec:app}
In the main text (Section~3), an estimate for the seeing noise is given. Here we describe the method used to obtain the value.

\begin{figure}[!tp]
\centering
\includegraphics[scale=0.5, clip=true, viewport=0.cm 0.0cm 18.cm 12.2cm]{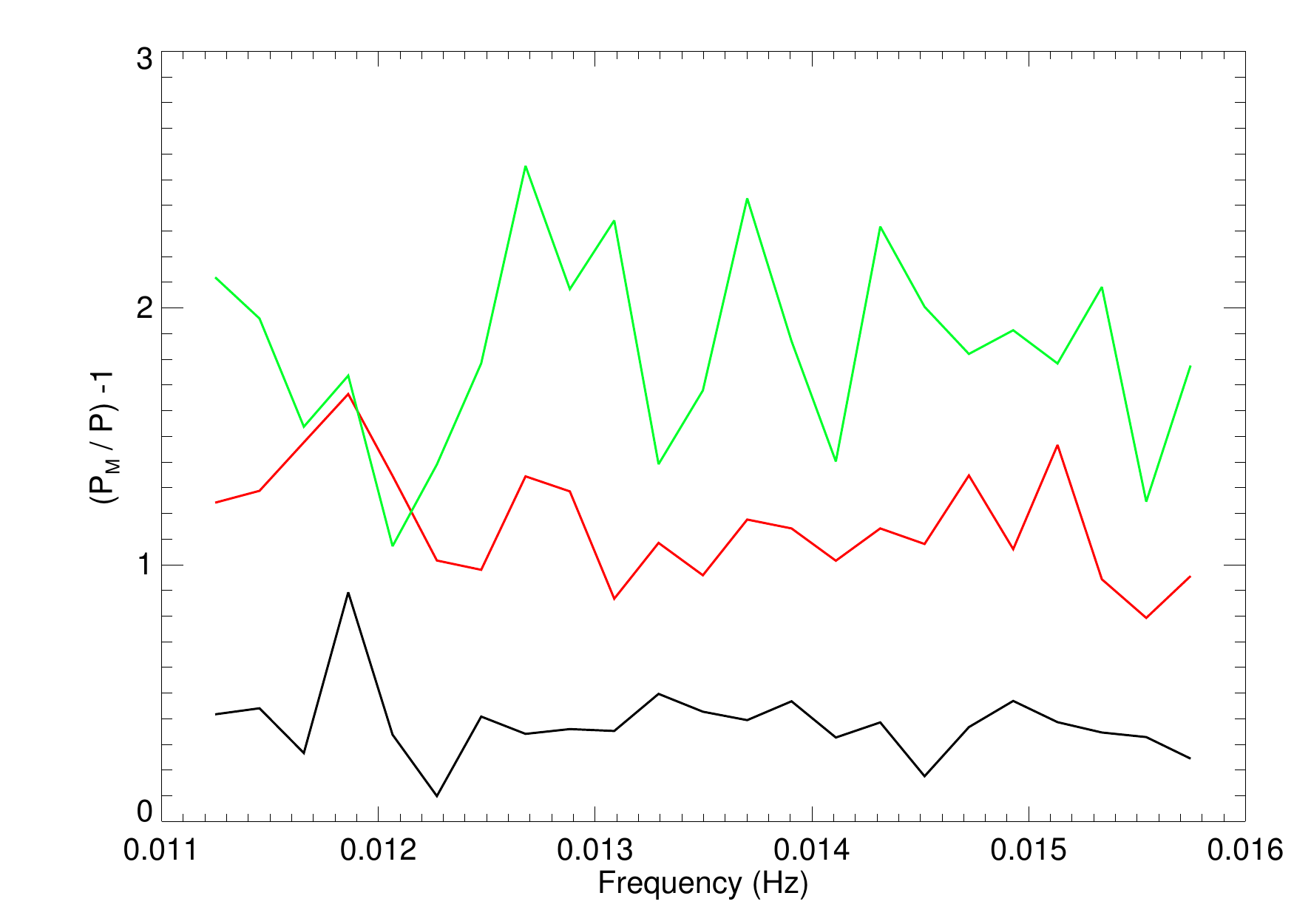} 
\caption{The ratio of the expected noise level to the measured noise level for the velocity power spectra. The black line corresponds to the open field region, the red is the quiet Sun region and the green is the active region.  The expected
noise levels used here are determined excluding the influence of seeing jitter. }\label{fig:noise_diff}
\end{figure}

Initially, the $\delta v$ for each pixel in each time frame is calculated from the contributions of photon, background and read noise. For each pixel, the uncertainty is then averaged over all time frames and the average uncertainty is used to generate a white noise time-series that is the same length as the data set. The power spectra of the noise is calculated for each pixel and averaged over the same 
spatial regions used for the velocity power spectra in Section~\ref{sec:av_pow}. As to be expected, the resultant noise power spectra is approximately flat. The spectra are fit with a linear function and the results are shown as the solid horizontal lines in Figure~\ref{fig:av_spec}. However, the estimated contribution of the uncertainties to the power is substantially less than the measured noise levels. To demonstrate this, the ratio of the power of the 
observed noise, $P_M$, to the expected power of the uncertainties, $P$, minus one, is shown in Figure~\ref{fig:noise_diff}. It can be clearly seen the level of 
underestimation varies between the regions, hence missing contribution is not a constant across the field of view.

The uncertainty (or noise) due to seeing conditions would have a dependence on the region under consideration due to the gradient of intensity differing between, say, an inhomogeneous active region and a more homogeneous coronal hole (see Eq.~\ref{eq:seeing}). In order to assess the affect of the seeing uncertainty, an estimate of $\sigma_z$ is needed. 

\begin{figure}[!tp]
\centering
\includegraphics[scale=0.65, clip=true, viewport=4.cm 0.0cm 18.cm 12.cm]{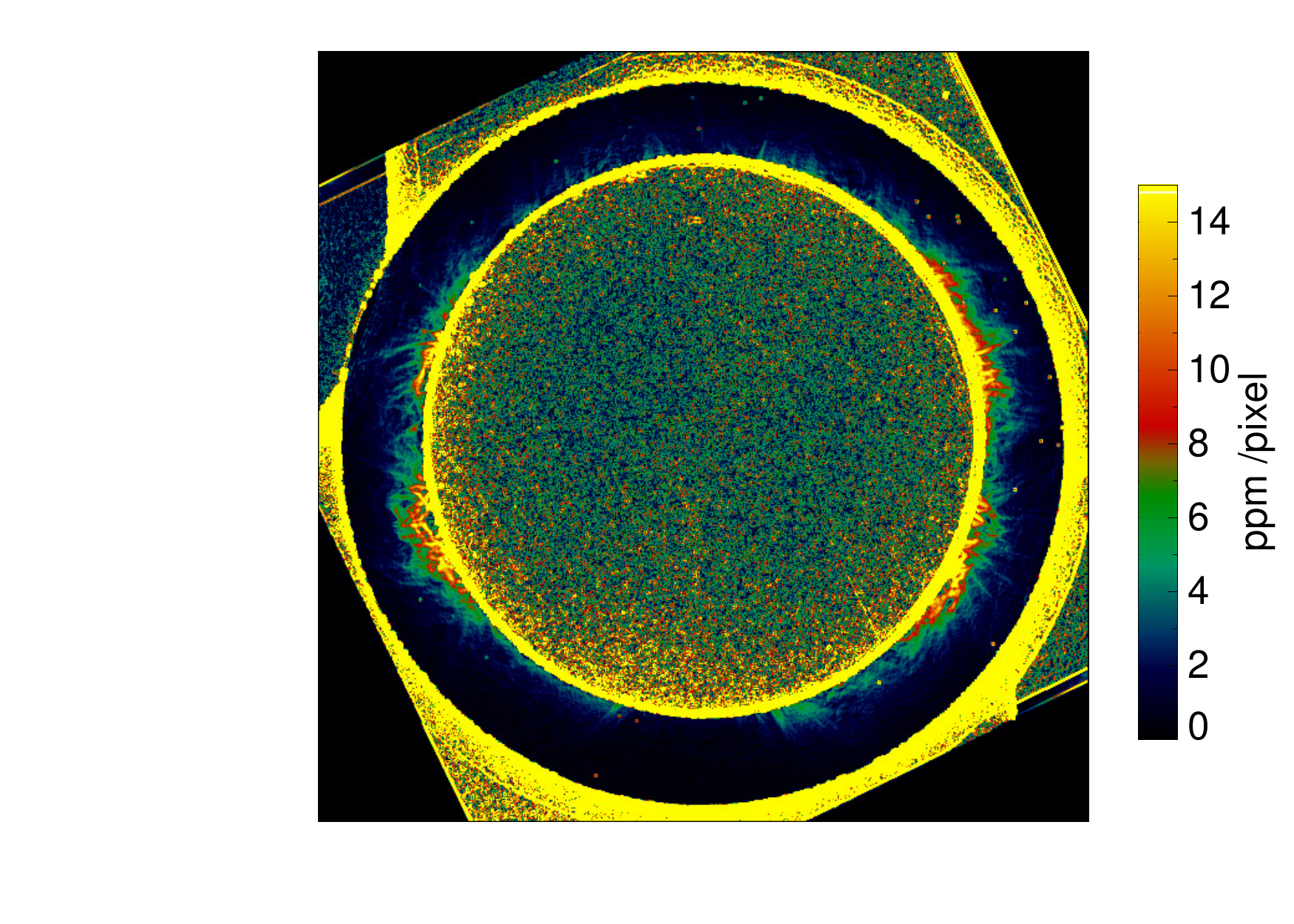} 
\caption{Magnitude gradient image. An example of the gradient of the intensity calculated using a Sobel filter.}\label{fig:grad}
\end{figure}

In the following, we demonstrate that it is possible to estimate the value analytically using the ratio $P_M/P$. As inferred from 
Eq.~(\ref{eq:gen_sum}), the error on velocity is given by,
\begin{equation}\label{eq:dv}
\delta v^2=\left(\frac{\partial v}{\partial \io}\delta \io\right)^2+\left(\frac{\partial v}{\partial \itw}\delta \itw\right)^2+\left(\frac{\partial v}{\partial 
\ith}\delta \ith\right).
\end{equation}
The partial derivatives are independent of the uncertainties so they can be written as constants, e.g., $A, B, C$. Additionally, the assumption is made that 
the uncertainties on each of the original intensity measurements is similar, i.e., $\delta \io\approx\delta \itw\approx\delta \ith$, hence, Eq.~(\ref{eq:dv}) 
can be written
\begin{equation}
\delta v^2=(A+B+C)\delta I^2.
\end{equation}
Now, taking the ratio of the estimated noise in the power, $\propto\delta v^2$, and the measured noise in the power spectra, $\propto\delta v_M^2$ (with associated uncertainties $\delta I_M$), gives the following relation
\begin{equation}\label{eq:im}
\delta I_M^2=\delta I^2\frac{\delta v_M^2}{\delta v^2},
\end{equation}
where $\delta I^2$, ${\delta v_M^2}$ and ${\delta v^2}$ are known. The $\delta I_M$ has contributions from the same uncertainties as $\delta I$ 
plus the addition of the unaccounted for noise sources, which is assumed to be only the seeing noise at present. Then, writing the measured uncertainty in intensity as
\begin{equation}
\delta I_M^2=\delta I^2+\sigma_{seeing}^2,
\end{equation}
and substituting into Eq.~(\ref{eq:im}) and rearranging gives
\begin{equation}
\sigma_{seeing}^2=\left(\frac{d I}{d z}\right)^2\sigma_z^2=\delta I^2\left(\frac{\delta v_M^2}{\delta v^2}-1\right).
\end{equation}
And finally, substituting in power for velocity squared we arrive at
\begin{equation}\label{eq:see_final}
\sigma_{seeing}^2=\left(\frac{d I}{d z}\right)^2\sigma_z^2=\delta I^2\left(\frac{P_M}{P}-1\right).
\end{equation}
Note, this equation only applies to the range of frequencies of the power spectra in Figure~\ref{fig:av_spec} in which the noise dominates. Lastly, an estimate 
of the gradient of the intensity for each pixel is required. To obtain this, a Sobel gradient operator is applied to an intensity image, which provides the 
magnitude gradient image (\citealp{GONWOO}), as shown in Figure~\ref{fig:grad}. It is clear that the intensity gradients are largest in 
the active region and least in the open field region. Now, using the noise dominated section of the velocity power spectra, i.e., from 10-16~mHz, an estimate for the $\sigma_{seeing}$ for each region is sought, and an average value of $0.1$~pixels is obtained. The 
corresponding uncertainty estimates demonstrate a much better agreement with the measured noise level (dashed horizontal lines in 
Figure~\ref{fig:av_spec}). 

\begin{thebibliography}{98}


\bibitem[{{Banerjee} {et~al.}(2011){Banerjee}, {Gupta}, \&
  {Teriaca}}]{BANetal2011}
{Banerjee}, D., {Gupta}, G.~R., \& {Teriaca}, L. 2011, \ssr, 158, 267

\bibitem[{{Banerjee} {et~al.}(2009){Banerjee}, {Teriaca}, {Gupta}, {Imada},
  {Stenborg}, \& {Solanki}}]{BANetal2009}
{Banerjee}, D., {Teriaca}, L., {Gupta}, G.~R., {et~al.} 2009, \aap, 499, L29

\bibitem[{{Bavassano} {et~al.}(1982){Bavassano}, {Dobrowolny}, {Mariani}, \&
  {Ness}}]{BAVetal1982}
{Bavassano}, B., {Dobrowolny}, M., {Mariani}, F., \& {Ness}, N.~F. 1982, \jgr,
  87, 3617

\bibitem[{{Berger} \& {Title}(1996)}]{BERTIT1996}
{Berger}, T.~E. \& {Title}, A.~M. 1996, \apj, 463, 365

\bibitem[{{Bethge} {et~al.}(2016){Bethge}, {Binay Karak}, {Tian}, {McIntosh},
  {Tomczyk}, {Threlfall}, \& {Sitongia}}]{BETetal2014}
{Bethge}, C., {Binay Karak}, B., {Tian}, H., {et~al.} 2016, In Prep

\bibitem[{{Bogdan} {et~al.}(2003){Bogdan}, {Carlsson}, {Hansteen}, {McMurry},
  {Rosenthal}, {Johnson}, {Petty-Powell}, {Zita}, {Stein}, {McIntosh}, \&
  {Nordlund}}]{BOGetal2003}
{Bogdan}, T.~J., {Carlsson}, M., {Hansteen}, V.~H., {et~al.} 2003, \apj, 599,
  626

\bibitem[{{Brooks} {et~al.}(2013){Brooks}, {Warren}, {Ugarte-Urra}, \&
  {Winebarger}}]{BROetal2013}
{Brooks}, D.~H., {Warren}, H.~P., {Ugarte-Urra}, I., \& {Winebarger}, A.~R.
  2013, ApJL, 772, L19

\bibitem[{{Bruno} \& {Carbone}(2005)}]{BRUCAR2005}
{Bruno}, R. \& {Carbone}, V. 2005, Living Reviews in Solar Physics, 2, 4

\bibitem[{{Cally}(2011)}]{CAL2011}
{Cally}, P.~S. 2011, in Astronomical Society of India Conference Series,
  Vol.~2, Astronomical Society of India Conference Series, 221--227

\bibitem[{{Cally} \& {Goossens}(20008)}]{CALGOO2008}
{Cally}, P.~S. \& {Goossens}, M. 20008, \solphys, 251, 251

\bibitem[{{Cally} \& {Hansen}(2011)}]{CALHAN2011}
{Cally}, P.~S. \& {Hansen}, S.~C. 2011, \apj, 738, 119

\bibitem[{{Chitta} {et~al.}(2012){Chitta}, {van Ballegooijen}, {Rouppe van der
  Voort}, {DeLuca}, \& {Kariyappa}}]{CHITetal2012}
{Chitta}, L.~P., {van Ballegooijen}, A.~A., {Rouppe van der Voort}, L.,
  {DeLuca}, E.~E., \& {Kariyappa}, R. 2012, \apj, 752, 48

\bibitem[{{Close} {et~al.}(2004){Close}, {Parnell}, {Longcope}, \&
  {Priest}}]{CLOetal2004}
{Close}, R.~M., {Parnell}, C.~E., {Longcope}, D.~W., \& {Priest}, E.~R. 2004,
  ApJL, 612, L81

\bibitem[{{Cranmer}(2012)}]{CRA2012}
{Cranmer}, S.~R. 2012, \ssr, 172, 145

\bibitem[{{Cranmer} \& {van Ballegooijen}(2005)}]{CRAVAN2005}
{Cranmer}, S.~R. \& {van Ballegooijen}, A.~A. 2005, \apjs, 156, 265

\bibitem[{{De Moortel}(2009)}]{DEM2009}
{De Moortel}, I. 2009, Space Science Reviews, 149, 65

\bibitem[{{De Moortel} {et~al.}(2014){De Moortel}, {McIntosh}, {Threlfall},
  {Bethge}, \& {Liu}}]{DEMetal2014}
{De Moortel}, I., {McIntosh}, S.~W., {Threlfall}, J., {Bethge}, C., \& {Liu},
  J. 2014, \apj, 782, L34

\bibitem[{{De Moortel} \& {Pascoe}(2012)}]{DEMPAS2012}
{De Moortel}, I. \& {Pascoe}, D.~J. 2012, \apj, 746, 31

\bibitem[{{De Pontieu} {et~al.}(2004){De Pontieu}, {Erd{\'e}lyi}, \&
  {James}}]{DEPetal2004}
{De Pontieu}, B., {Erd{\'e}lyi}, R., \& {James}, S.~P. 2004, \nat, 430, 536

\bibitem[{{De Pontieu} {et~al.}(2007{\natexlab{a}}){De Pontieu}, {Hansteen},
  {Rouppe van der Voort}, {van Noort}, \& {Carlsson}}]{DEPetal2007c}
{De Pontieu}, B., {Hansteen}, V.~H., {Rouppe van der Voort}, L., {van Noort},
  M., \& {Carlsson}, M. 2007{\natexlab{a}}, \apj, 655, 624

\bibitem[{{De Pontieu} {et~al.}(2007{\natexlab{b}}){De Pontieu}, {McIntosh},
  {Carlsson}, {Hansteen}, {Tarbell}, {Schrijver}, {Title}, {Shine}, {Tsuneta},
  {Katsukawa}, {Ichimoto}, {Suematsu}, {Shimizu}, \& {Nagata}}]{DEPetal2007}
{De Pontieu}, B., {McIntosh}, S.~W., {Carlsson}, M., {et~al.}
  2007{\natexlab{b}}, Science, 318, 1574

\bibitem[{{DeForest} \& {Gurman}(1998)}]{DEFGRU1998}
{DeForest}, C.~E. \& {Gurman}, J.~B. 1998, ApJL, 501, L217

\bibitem[{{DeForest} {et~al.}(2014){DeForest}, {Howard}, \&
  {McComas}}]{DEFetal2014}
{DeForest}, C.~E., {Howard}, T.~A., \& {McComas}, D.~J. 2014, \apj, 787, 124

\bibitem[{{Dowdy} {et~al.}(1986){Dowdy}, {Rabin}, \& {Moore}}]{DOWetal1986}
{Dowdy}, Jr., J.~F., {Rabin}, D., \& {Moore}, R.~L. 1986, \solphys, 105, 35

\bibitem[{{Edwin} \& {Roberts}(1983)}]{EDWROB1983}
{Edwin}, P.~M. \& {Roberts}, B. 1983, \solphys, 88, 179

\bibitem[{{Erd{\'e}lyi} \& {Ballai}(2007)}]{ERDBAL2007}
{Erd{\'e}lyi}, R. \& {Ballai}, I. 2007, Astronomische Nachrichten, 328, 726

\bibitem[{{Erd{\'e}lyi} \& {Taroyan}(2008)}]{ERDTAR2008}
{Erd{\'e}lyi}, R. \& {Taroyan}, Y. 2008, \aap, 489, L49

\bibitem[{{Fedun} {et~al.}(2009){Fedun}, {Erd{\'e}lyi}, \&
  {Shelyag}}]{FEDetal2009}
{Fedun}, V., {Erd{\'e}lyi}, R., \& {Shelyag}, S. 2009, \solphys, 258, 219

\bibitem[{{Fedun} {et~al.}(2011){Fedun}, {Shelyag}, \&
  {Erd{\'e}lyi}}]{FEDetal2011}
{Fedun}, V., {Shelyag}, S., \& {Erd{\'e}lyi}, R. 2011, \apj, 727, 17

\bibitem[{{Flower} \& {Pineau des Forets}(1973)}]{FLOPIN1973}
{Flower}, D.~R. \& {Pineau des Forets}, G. 1973, \aap, 24, 181

\bibitem[{{Gabriel}(1976)}]{GAB1976}
{Gabriel}, A.~H. 1976, Royal Society of London Philosophical Transactions
  Series A, 281, 339

\bibitem[{{Gascoyne} {et~al.}(2014){Gascoyne}, {Jain}, \&
  {Hindman}}]{GASetal2014}
{Gascoyne}, A., {Jain}, R., \& {Hindman}, B.~W. 2014, \apj, 789, 109

\bibitem[{{Goldstein} {et~al.}(1995){Goldstein}, {Smith}, {Balogh}, {Horbury},
  {Goldstein}, \& {Roberts}}]{GOLetal1995}
{Goldstein}, B.~E., {Smith}, E.~J., {Balogh}, A., {et~al.} 1995, \grl, 22, 3393

\bibitem[{{Gonzalez} \& {Woods}(2002)}]{GONWOO}
{Gonzalez}, R.~C. \& {Woods}, R.~E. 2002, Digital image processing, 2nd edn.
  (Upper Saddle River, NJ: Prentice Hall)

\bibitem[{{Goossens} {et~al.}(2009){Goossens}, {Terradas}, {Andries},
  {Arregui}, \& {Ballester}}]{GOOetal2009}
{Goossens}, M., {Terradas}, J., {Andries}, J., {Arregui}, I., \& {Ballester},
  J.~L. 2009, \aap, 503, 213
  
  \bibitem[{{Goossens} {et~al.}(2012){Goossens}, {Andries}, {Soler}, {Van Doorsselaere}
  {Arregui}, \& {Terradas}}]{GOOetal2012}
{Goossens}, M., {Andries}, {Soler}, R., {Van Doorsselaere}, T.,J., {Arregui}, I., \& {Terradas}, J.
   2012, \apj, 753, 111

\bibitem[{{Goossens} {et~al.}(2013){Goossens}, {Van Doorsselaere}, {Soler}, \&
  {Verth}}]{GOOetal2013}
{Goossens}, M., {Van Doorsselaere}, T., {Soler}, R., \& {Verth}, G. 2013, \apj,
  768, 191

\bibitem[{{Gupta}(2014)}]{GUP2014}
{Gupta}, G.~R. 2014, \aap, 568, A96

\bibitem[{{Hansen} \& {Cally}(2012)}]{HANCAL2012}
{Hansen}, S.~C. \& {Cally}, P.~S. 2012, \apj, 751, 31

\bibitem[{{Hansteen} \& {Velli}(2012)}]{HANVEL2012}
{Hansteen}, V.~H. \& {Velli}, M. 2012, \ssr, 172, 89

\bibitem[{{He} {et~al.}(2009){He}, {Marsch}, {Tu} \&
  {Tian}}]{HEetal2009}
{He}, J.~S., {Marsch}, E., {Tu}, C., \& {Tian}, H. 2009, ApJL, 705, L217

\bibitem[{{Hillier} {et~al.}(2013){Hillier}, {Morton}, \&
  {Erd\'elyi}}]{HILetal2013}
{Hillier}, A., {Morton}, R.~J., \& {Erd\'elyi}, R. 2013, ApJL, 779, L16

\bibitem[{{Hollweg}(1978)}]{HOL1978}
{Hollweg}, J.~V. 1978, \solphys, 56, 305

\bibitem[{{Jain} {et~al.}(2011){Jain}, {Gascoyne}, \& {Hindman}}]{JAIetal2011}
{Jain}, R., {Gascoyne}, A., \& {Hindman}, B.~W. 2011, \mnras, 415, 1276

\bibitem[{{Jess} {et~al.}(2015){Jess}, {Morton}, {Verth}, {Fedun}, {Grant}, \&
  {Giagkiozis}}]{JESetal2015}
{Jess}, D.~B., {Morton}, R.~J., {Verth}, G., {et~al.} 2015, \ssr, 190, 103

\bibitem[{{Keys} {et~al.}(2011){Keys}, {Mathioudakis}, {Jess}, {Shelyag},
  {Crockett}, {Christian}, \& {Keenan}}]{KEYetal2011}
{Keys}, P.~H., {Mathioudakis}, M., {Jess}, D.~B., {et~al.} 2011, ApJL, 740,
  L40

\bibitem[{{Khomenko} \& {Cally}(2012)}]{KHOCAL2012}
{Khomenko}, E. \& {Cally}, P.~S. 2012, \apj, 746, 68

\bibitem[{{Khomenko} \& {Collados}(2006)}]{KHOCOL2006}
{Khomenko}, E. \& {Collados}, M. 2006, \apj, 653, 739

\bibitem[{{Khomenko} {et~al.}(2008){Khomenko}, {Collados}, \&
  {Felipe}}]{KHOetal2008}
{Khomenko}, E., {Collados}, M., \& {Felipe}, T. 2008, \solphys, 251, 589

\bibitem[{{Klimchuk}(2006)}]{KLI2006}
{Klimchuk}, J.~A. 2006, \solphys, 234, 41

\bibitem[{{Krishna Prasad} {et~al.}(2014){Krishna Prasad}, {Banerjee}, \& {Van
  Doorsselaere}}]{KRIetal2014}
{Krishna Prasad}, S., {Banerjee}, D., \& {Van Doorsselaere}, T. 2014, \apj,
  789, 118

\bibitem[{{Landi} {et~al.}(2012){Landi}, {Del Zanna}, {Young}, {Dere}, \&
  {Mason}}]{LANetal2012}
{Landi}, E., {Del Zanna}, G., {Young}, P.~R., {Dere}, K.~P., \& {Mason}, H.~E.
  2012, \apj, 744, 99

\bibitem[{{Lemen} {et~al.}(2012){Lemen}, {Title}, {Akin}, {Boerner}, {Chou},
  {Drake}, {Duncan}, {Edwards}, {Friedlaender}, {Heyman}, {Hurlburt}, {Katz},
  {Kushner}, {Levay}, {Lindgren}, {Mathur}, {McFeaters}, {Mitchell}, {Rehse},
  {Schrijver}, {Springer}, {Stern}, {Tarbell}, {Wuelser}, {Wolfson}, {Yanari},
  {Bookbinder}, {Cheimets}, {Caldwell}, {Deluca}, {Gates}, {Golub}, {Park},
  {Podgorski}, {Bush}, {Scherrer}, {Gummin}, {Smith}, {Auker}, {Jerram},
  {Pool}, {Soufli}, {Windt}, {Beardsley}, {Clapp}, {Lang}, \&
  {Waltham}}]{LEMetal2012}
{Lemen}, J.~R., {Title}, A.~M., {Akin}, D.~J., {et~al.} 2012, \solphys, 275, 17

\bibitem[{Lilliefors(1969)}]{LIL1967}
Lilliefors, H. 1969, J AMER STATIST ASSN, 62, 399

\bibitem[{{Markwardt}(2009)}]{MAR2009}
{Markwardt}, C.~B. 2009, in ASPCS, Vol. 411, Astronomical Data Analysis
  Software and Systems XVIII, ed. D.~A. {Bohlender}, D.~{Durand}, \&
  P.~{Dowler}, 251

\bibitem[{{Matthaeus} \& {Velli}(2011)}]{MATVEL2011}
{Matthaeus}, W.~H. \& {Velli}, M. 2011, \ssr, 160, 145

\bibitem[{{McIntosh} \& {De Pontieu}(2012)}]{MCIDEP2012}
{McIntosh}, S.~W. \& {De Pontieu}, B. 2012, \apj, 761, 138

\bibitem[{{McIntosh} {et~al.}(2011){McIntosh}, {de Pontieu}, {Carlsson},
  {Hansteen}, {Boerner}, \& {Goossens}}]{MCIetal2011}
{McIntosh}, S.~W., {de Pontieu}, B., {Carlsson}, M., {et~al.} 2011, \nat, 475,
  477

\bibitem[{{Morgan} \& {Druckm{\"u}ller}(2014)}]{MORDRU2014}
{Morgan}, H. \& {Druckm{\"u}ller}, M. 2014, \solphys, 289, 2945

\bibitem[{{Morton} \& {McLaughlin}(2013)}]{MORMCL2013}
{Morton}, R.~J. \& {McLaughlin}, J.~A. 2013, \aap, 553, 10

\bibitem[{{Morton} \& {McLaughlin}(2014)}]{MORMCL2014}
{Morton}, R.~J. \& {McLaughlin}, J.~A. 2014, \apj, 789, 105

\bibitem[{{Morton} {et~al.}(2015){Morton}, {Tomczyk}, \& {Pinto}}]{MORetal2015}
{Morton}, R.~J., {Tomczyk}, S., \& {Pinto}, R. 2015, Nature Comms., 6, 7813

\bibitem[{{Morton} {et~al.}(2013){Morton}, {Verth}, {Fedun}, {Shelyag}, \&
  {Erd\'{e}lyi}}]{MORetal2013}
{Morton}, R.~J., {Verth}, G., {Fedun}, V., {Shelyag}, S., \& {Erd\'{e}lyi}, R.
  2013, \apj, 768, 17

\bibitem[{{Morton} {et~al.}(2014){Morton}, {Verth}, {Hillier}, \&
  {Erd\'elyi}}]{MORetal2013b}
{Morton}, R.~J., {Verth}, G., {Hillier}, A., \& {Erd\'elyi}, R. 2014, ApJ, 784,
  29

\bibitem[{{Morton} {et~al.}(2012){Morton}, {Verth}, {Jess}, {Kuridze},
  {Ruderman}, {Mathioudakis}, \& {Erd{\'e}lyi}}]{MORetal2012c}
{Morton}, R.~J., {Verth}, G., {Jess}, D.~B., {et~al.} 2012, Nat. Commun., 3,
  1315

\bibitem[{{Narain} \& {Ulmschneider}(1996)}]{NARULM1996}
{Narain}, U. \& {Ulmschneider}, P. 1996, \ssr, 75, 453

\bibitem[{{Nistic{\`o}} {et~al.}(2013){Nistic{\`o}}, {Nakariakov}, \&
  {Verwichte}}]{NISetal2013}
{Nistic{\`o}}, G., {Nakariakov}, V.~M., \& {Verwichte}, E. 2013, \aap, 552, A57

\bibitem[{{Ofman} {et~al.}(1997){Ofman}, {Romoli}, {Poletto}, {Noci}, \&
  {Kohl}}]{OFMetal1997}
{Ofman}, L., {Romoli}, M., {Poletto}, G., {Noci}, G., \& {Kohl}, J.~L. 1997,
  ApJL, 491, L111

\bibitem[{{Okamoto} {et~al.}(2007){Okamoto}, {Tsuneta}, {Berger}, {Ichimoto},
  {Katsukawa}, {Lites}, {Nagata}, {Shibata}, {Shimizu}, {Shine}, {Suematsu},
  {Tarbell}, \& {Title}}]{OKAetal2007}
{Okamoto}, T.~J., {Tsuneta}, S., {Berger}, T.~E., {et~al.} 2007, Science, 318,
  1577

\bibitem[{Olsen(1993)}]{OLS1993}
Olsen, S.~I. 1993, CVGIP: Graphical Models and Image Processing, 55, 319

\bibitem[{{Osterbrock}(1961)}]{OST1961}
{Osterbrock}, D.~E. 1961, \apj, 134, 347

\bibitem[{{Parnell} \& {De Moortel}(2012)}]{PARDEM2012}
{Parnell}, C.~E. \& {De Moortel}, I. 2012, Royal Society of London
  Philosophical Transactions Series A, 370, 3217

\bibitem[{{Pereira} {et~al.}(2012){Pereira}, {De Pontieu}, \&
  {Carlsson}}]{PERetal2012}
{Pereira}, T.~M., {De Pontieu}, B., \& {Carlsson}, M. 2012, \apj, 759, 16

\bibitem[{{Pesnell} {et~al.}(2012){Pesnell}, {Thompson}, \&
  {Chamberlin}}]{PESetal2012}
{Pesnell}, W.~D., {Thompson}, B.~J., \& {Chamberlin}, P.~C. 2012, \solphys,
  275, 3

\bibitem[{{Peter}(2001)}]{PET2001}
{Peter}, H. 2001, \aap, 374, 1108

\bibitem[{{Petrosyan} {et~al.}(2010){Petrosyan}, {Balogh}, {Goldstein},
  {L{\'e}orat}, {Marsch}, {Petrovay}, {Roberts}, {von Steiger}, \&
  {Vial}}]{PETetal2010}
{Petrosyan}, A., {Balogh}, A., {Goldstein}, M.~L., {et~al.} 2010, \ssr, 156,
  135

\bibitem[{{Roberts}(2010)}]{DROB2010}
{Roberts}, D.~A. 2010, \apj, 711, 1044

\bibitem[{{Scherrer} {et~al.}(2012){Scherrer}, {Schou}, {Bush}, {Kosovichev},
  {Bogart}, {Hoeksema}, {Liu}, {Duvall}, {Zhao}, {Title}, {Schrijver},
  {Tarbell}, \& {Tomczyk}}]{SCHetal2012}
{Scherrer}, P.~H., {Schou}, J., {Bush}, R.~I., {et~al.} 2012, \solphys, 275,
  207

\bibitem[{{Schrijver} \& {De Rosa}(2003)}]{SCHDER2003}
{Schrijver}, C.~J. \& {De Rosa}, M.~L. 2003, \solphys, 212, 165

\bibitem[{{Schunker} \& {Cally}(2006)}]{SCHCAL2006}
{Schunker}, H. \& {Cally}, P.~S. 2006, \mnras, 372, 551

\bibitem[{{Spruit}(1982)}]{SPR1982}
{Spruit}, H.~C. 1982, \solphys, 75, 3

\bibitem[{{Stangalini} {et~al.}(2013){Stangalini}, {Berrilli}, \&
  {Consolini}}]{STAetal2013}
{Stangalini}, M., {Berrilli}, F., \& {Consolini}, G. 2013, \aap, 559, A88

\bibitem[{{Stangalini} {et~al.}(2014){Stangalini}, {Consolini}, {Berrilli}, {De
  Michelis}, \& {Tozzi}}]{STAetal2014}
{Stangalini}, M., {Consolini}, G., {Berrilli}, F., {De Michelis}, P., \&
  {Tozzi}, R. 2014, \aap, 569, A102

\bibitem[{{Stangalini} {et~al.}(2015){Stangalini}, {Giannattasio}, \&
  {Jafarzadeh}}]{STAetal2015}
{Stangalini}, M., {Giannattasio}, F., \& {Jafarzadeh}, S. 2015, \aap, 577, A17

\bibitem[{{Starck} \& {Murtagh}(2006)}]{STAMUR2006}
{Starck}, J.-L. \& {Murtagh}, F. 2006, Astronomical Image and Data Analysis,
  2nd edn. (Springer-Verlag Berlin Heidelberg)

\bibitem[{{Threlfall} {et~al.}(2013){Threlfall}, {De Moortel}, {McIntosh}, \&
  {Bethge}}]{THRetal2013}
{Threlfall}, J., {De Moortel}, I., {McIntosh}, S.~W., \& {Bethge}, C. 2013,
  \aap, 556, A124

\bibitem[{{Thurgood} {et~al.}(2014){Thurgood}, {Morton}, \&
  {McLaughlin}}]{THUetal2014}
{Thurgood}, J.~O., {Morton}, R.~J., \& {McLaughlin}, J.~A. 2014, ApJL, 790, L2

\bibitem[{{Tian} {et~al.}(2013){Tian}, {Tomczyk}, {McIntosh}, {Bethge}, {de
  Toma}, \& {Gibson}}]{TIAetal2013}
{Tian}, H., {Tomczyk}, S., {McIntosh}, S.~W., {et~al.} 2013, \solphys, 288, 637

\bibitem[{{Title} {et~al.}(1989){Title}, {Tarbell}, {Topka}, {Ferguson},
  {Shine}, \& {SOUP Team}}]{TITetal1989}
{Title}, A.~M., {Tarbell}, T.~D., {Topka}, K.~P., {et~al.} 1989, \apj, 336, 475

\bibitem[{{Tomczyk} {et~al.}(2008){Tomczyk}, {Card}, {Darnell}, {Elmore},
  {Lull}, {Nelson}, {Streander}, {Burkepile}, {Casini}, \&
  {Judge}}]{TOMetal2008}
{Tomczyk}, S., {Card}, G.~L., {Darnell}, T., {et~al.} 2008, \solphys, 247, 411

\bibitem[{{Tomczyk} \& {McIntosh}(2009)}]{TOMMCI2009}
{Tomczyk}, S. \& {McIntosh}, S.~W. 2009, \apj, 697, 1384

\bibitem[{{Tomczyk} {et~al.}(2007){Tomczyk}, {McIntosh}, {Keil}, {Judge},
  {Schad}, {Seeley}, \& {Edmondson}}]{TOMetal2007}
{Tomczyk}, S., {McIntosh}, S.~W., {Keil}, S.~L., {et~al.} 2007, Science, 317,
  1192

\bibitem[{{van Ballegooijen} {et~al.}(2011){van Ballegooijen}, {Asgari-Targhi},
  {Cranmer}, \& {DeLuca}}]{VANBALLetal2011}
{van Ballegooijen}, A.~A., {Asgari-Targhi}, M., {Cranmer}, S.~R., \& {DeLuca},
  E.~E. 2011, \apj, 736, 3

\bibitem[{{van Ballegooijen} {et~al.}(1998){van Ballegooijen}, {Nisenson},
  {Noyes}, {L{\"o}fdahl}, {Stein}, {Nordlund}, \&
  {Krishnakumar}}]{VANBALLetal1998}
{van Ballegooijen}, A.~A., {Nisenson}, P., {Noyes}, R.~W., {et~al.} 1998, \apj,
  509, 435

\bibitem[{{Van Doorsselaere} {et~al.}(2014){Van Doorsselaere}, {Gijsen},
  {Andries}, \& {Verth}}]{VANetal2014}
{Van Doorsselaere}, T., {Gijsen}, S.~E., {Andries}, J., \& {Verth}, G. 2014,
  \apj, 795, 18

\bibitem[{{van Doorsselaere} {et~al.}(2008){van Doorsselaere}, {Nakariakov},
  {Young}, \& {Verwichte}}]{VANetal2008c}
{van Doorsselaere}, T., {Nakariakov}, V.~M., {Young}, P.~R., \& {Verwichte}, E.
  2008, \aap, 487, L17

\bibitem[{{Verdini} {et~al.}(2012){Verdini}, {Grappin}, {Pinto}, \&
  {Velli}}]{VERDetal2012}
{Verdini}, A., {Grappin}, R., {Pinto}, R., \& {Velli}, M. 2012, ApJL, 750, L33

\bibitem[{{Verdini} \& {Velli}(2007)}]{VERVEL2007}
{Verdini}, A. \& {Velli}, M. 2007, \apj, 662, 669

\bibitem[{{Vigeesh} {et~al.}(2009){Vigeesh}, {Hasan}, \&
  {Steiner}}]{VIGetal2009}
{Vigeesh}, G., {Hasan}, S.~S., \& {Steiner}, O. 2009, \aap, 508, 951

\bibitem[{{Winebarger} {et~al.}(2014){Winebarger}, {Cirtain}, {Golub},
  {DeLuca}, {Savage}, {Alexander}, \& {Schuler}}]{WINetal2014}
{Winebarger}, A.~R., {Cirtain}, J., {Golub}, L., {et~al.} 2014, ApJL, 787, L10

\end{thebibliography}
\end{document}